\documentclass{siamltex}

\usepackage{amsmath}
\usepackage{amssymb}
\usepackage{graphicx}

\newtheorem{thm}{Theorem}

\title{Splay states in finite pulse-coupled networks of excitable neurons}

\author{M. Dipoppa\footnotemark[2]\ \footnotemark[3]\ \footnotemark[4]
 \and M. Krupa\footnotemark[2]\ \footnotemark[5] 
  \and A. Torcini\footnotemark[6]\ \footnotemark[7]\ \ \footnotemark[8]\ \ \footnotemark[9]
 \and B.~S. Gutkin\footnotemark[2]\ \footnotemark[3]\ \ \footnotemark[9]}
 
\begin{document}

 \maketitle
 
 \renewcommand{\thefootnote}{\fnsymbol{footnote}}
 
 \footnotetext[2]{Group for
Neural Theory, LNC, D\'epartement d'Etudes Cognitives, Ecole Normale Sup\'erieure, 29 rue d'Ulm 75005 Paris, France ({\tt mario.dipoppa@ens.fr}).}
\footnotetext[3]{Laboratoire de Neurosciences Cognitives, INSERM U960, 29 rue d'Ulm 75005 Paris, France.}
   \footnotetext[4]{Universit\'{e} Pierre et Marie Curie, 4 place Jussieu 75005 Paris, France.}
    \footnotetext[5]{ Donders Institute for Brain, Cognition and Behaviour, Department of Medical Physics and Biophysics, Radboud Universiteit Nijmegen, Geert Grooteplein 21, NL 6525 EZ Nijmegen, The Netherlands.}
  \footnotetext[6]{CNR - Istituto dei Sistemi Complessi, via Madonna del Piano 10, I-50019 Sesto Fiorentino, Italy.}
\footnotetext[7]{INFN Sez. Firenze, via Sansone, 1 - I-50019 Sesto Fiorentino, Italy.}
 \footnotetext[8]{Centro Interdipartimentale per lo Studio delle Dinamiche Complesse, via Sansone, 1 - I-50019 Sesto Fiorentino, Italy.}
   \footnotetext[9]{Joint senior autorship.}
\renewcommand{\thefootnote}{\arabic{footnote}}

 \maketitle

\begin{abstract}
The emergence and stability of splay states is studied in fully
coupled finite networks of $N$ excitable quadratic integrate-and-fire
neurons, connected via synapses modeled as  pulses of finite amplitude
and duration. For such synapses, by introducing two distinct types of synaptic
events (pulse emission and termination), we were able to 
write down an exact event-driven map for the system and to evaluate the
splay state solutions. For $M$ overlapping post synaptic potentials 
the linear stability analysis of the splay state should take
in account, besides the actual values of the membrane potentials, also
the firing times associated to the $M$ previous pulse emissions.
As a matter of fact, it was possible, by introducing $M$ complementary variables, 
to rephrase the evolution of the network as an event-driven map 
and to derive an  analytic expression for the Floquet
spectrum. We find that, independently of $M$, the splay state is
marginally stable with $N-2$ neutral directions. Furthermore, 
we have identified a family of periodic solutions 
surrounding the splay state and sharing the same neutral stability directions.
In the limit of $\delta$-pulses, it is still possible to derive
an event-driven formulation for the dynamics, however the number of neutrally stable
directions, associated to the splay state, becomes $N$.
Finally, we prove a link between the results for our system and a previous theory
[{\it Watanabe and Strogatz, Physica D, 74 (1994), pp. 197- 253}]
developed for networks of phase oscillators with sinusoidal coupling. 
\end{abstract}



	


\pagestyle{myheadings}
\thispagestyle{plain}
\markboth{DIPOPPA, KRUPA, TORCINI, AND GUTKIN}{SPLAY STATES IN NETWORKS OF EXCITABLE NEURONS}

 \section{Introduction}
The dynamics of networks made up of many elements with a high degree of 
connectivity is often studied in the infinite size limit,
which allows to apply approaches borrowed from statistical physics.
In particular, for globally coupled neural networks this amounts 
to finding the distribution of the membrane potentials 
satisfying a Fokker-Planck equation with specific boundary conditions corresponding
to the spike emission and reset of the neurons \cite{amit,abbott}.
In contrast, general techniques to deal with the dynamics of finite size ensembles are not 
yet fully developed, not even for the analysis of the linear stability
of periodic solutions.

In this paper we investigate the stability of {\it splay states} (also known as
antiphase states or ``ponies on a merry-go-round")~\cite{hadley,krupa}. 
In a splay state all the $N$ elements follow the same periodic 
dynamics $x(t)$ ($x(t+ N \cdot T)=x(t)$) but with different time shifts evenly 
distributed at regular intervals $\Delta T = k T$, with $k=1,\dots,N$.
Experimental observations of splay states have been reported in  multimode 
laser systems \cite{wiese90} and electronic circuits \cite{ashw90}.
Numerical and theoretical analyses have been devoted to
splay states in Josephson junction arrays \cite{hadley,nicols,stro2,krupa}, globally coupled 
Ginzburg-Landau equations \cite{hakim92}, globally coupled laser models \cite{rappel94},
traffic models \cite{seidel}, and pulse-coupled neuronal networks \cite{abbott}. 
In the latter context, splay states have been usually investigated for
leaky-integrate-and-fire (LIF) neurons and in general for neuronal
models which can be assimilated to phase oscillators (rotators)
\cite{abbott,vvres,Zillmer_et_al_2007}.
The first detailed stability analysis of LIF neuron oscillators was performed by developing
a mean-field approach in the infinite network limit \cite{abbott,vvres}. Finite size stability analysis for 
supra-threshold neurons, namely for LIF~\cite{Zillmer_et_al_2007} and for 
generalized neuronal models in the oscillatory regime~\cite{Calamai_et_al_2009}, have been more recently
developed, based on the linearization of a suitable Poincar\'e map.

The model analyzed in this paper is a a fully coupled network of excitable neurons, governed by
the quadratic integrate-and-fire equation (QIF). The QIF is the canonical model for
 type I neuronal excitability as it is the quadratic normal form for the saddle-node
  invariant cycle (SNIC)  bifurcation~\cite{erme}. The neurons are coupled with
   positive pulses, modeling excitatory synapses.  We focus our analysis on the persistent activity of the
network that is induced by the recurrent excitation and that co-exists with an inactive ground state. 

Analyzing this type of activity is of significant relevance to neuroscience.
 The bistable sustained activity has been shown to be the neuronal basis for 
 working memory \cite{funa, fuster}.
  Such self-sustained elevated activity has been recorded in delayed response tasks where the memory trace
   must be retained in order to generate appropriate responses. 
   Furthermore, so-called cortical up-states, observed during anesthesia and during sleep,
    are also considered to be generated by the intrinsic excitatory 
    synaptic connectivity with the constituent neurons being excitable
     (as opposed to intrinsic oscillators). 
There are also indications that these sustained up-states are largely asynchronous.
 In fact, theoretical studies
 have suggested that asynchrony is a requirement for stable maintenance of synaptically
  sustained neural activity \cite{laingchow}.

Furthermore, previous computational work proposed that perturbing the asynchronous 
structure of the sustained activity leads to its destabilization 
\cite{Gutkin_et_al_2001,compte}. It is thus important to determine 
specifically the stability and the structure of the asynchronous sustained activity. This item has been addressed in the infinite size limit within the mean-field 
approximation \cite{Hansel_Mato_2001,Hansel_Mato_2003} and the 
the role of asynchrony and synchrony in sustained neural activity
has been studied for a pair of neurons
\cite{Gutkin_et_al_2001}.
However sustained cortical activity appears to be generated by local circuits in the cortex, 
i.e. networks with a limited number of neurons. Hence in our work we seek
 to understand the stability of asynchronous activity self-sustained by a finite size network.
 
 In this paper, as already mentioned, we analyze the splay states, which represent
highly symmetric states.  
We perform an analytical linear stability analysis of the splay states
for finite size networks when the post-synaptic potentials (PSPs) are
modeled as square pulses of finite amplitude and duration.
We focus on fast excitatory synaptic coupling as a basic mechanism
to generate the reverberative self-sustained activity. This corresponds to AMPA
receptor-mediated glutamatergic synapses that have a typical decay-time constant of about $5$ msecs \cite{spruston}.
Traditionally such synapses are modeled as a double exponential function
(or an $\alpha$-function) with a finite rise-time and a decay-time governed by
the synaptic time-constant \cite{gerstner}.

Here we use a simpler version of this model: we
keep the idea of the characteristic synaptic time scale while leaving aside
the dynamics by modeling the synaptic currents as square pulse steps.
The advantage of such a minimal model is that it makes the network dynamics
tractable for our analysis, while giving us control over the synaptic duration.

In order to study the finite size network, we derive an event driven map for the
evolution of the membrane potentials
of the neurons, by introducing two kinds of synaptic events:
synaptic pulse emission and termination. This approach allows us to derive an
analytic, but implicit, expression for the splay state for two kinds of synaptic models:
step and $\delta$-pulses. Furthermore, the linear stability analysis requires the investigation of the
linearized dynamics of the model. It should be mentioned that
memory effects should be taken in account whenever the duration of the
post-synaptic potentials last sufficiently to lead to overlaps among the
emitted pulses. For $M$ overlapping pulses, the linearized dynamics can
be rewritten as an event-driven map by including $M$ additional 
variables.  This at variance with the usual approach, where
the memory effect due to the linear super-position of 
$\alpha$- or exponential pulses emitted in the past 
is taken in account by a self-consistent field~\cite{abbott}.
Finally, by employing the event-driven formulation we have analytically obtained the Floquet spectra 
associated to the splay state for step and $\delta$-pulses.

The paper is organized as follows.
In \S2 the model and the possible dynamical
regimes are introduced. The event-driven map for step and
$\delta$-pulses is derived in \S3, while the linear
stability analysis of splay states is performed in \S4
for step pulses and in \S5 for $\delta$-pulses.
\S6 is devoted to the description of other
periodic states observable in the present model.
Finally, in \S7 the results are summarized
and discussed. Analytical expression for the firing
rates of the splay states in small networks are
reported in \S A. 
Furthermore, in \S B we report
an analytical expression for
the splay state membrane potentials derived 
in the continuum limit.
\S C contains a
formal proof for our model, in the
case of non-overlapping pulses, that the Floquet
spectrum associated with the splay states contains 
$N-2$ marginally stable directions.

\section{Model and Dynamical Regimes} 
In this Section, we will introduce our model and the specific state which
is the main subject of investigation of our analysis, namely the {\it splay state}.
In particular, we consider a pulse-coupled fully connected excitatory network
made of quadratic integrate and fire neurons, whose dynamics is governed by the following equation
\begin{equation} 
\label{QIF}  
\tau\frac{dv_i}{dt}=v_i^2-1+I(t) \qquad i=1,\dots,N 
\end{equation} 
where the $n$-th spike is emitted at time $t_n$, once the
neuron reaches the threshold value $v_i(t^-_n)=\infty$; afterwards it is
immediately reset to the value $v_i(t^+_n)=-\infty$. 
For a constant synaptic current $I < 1$, 
the neuron has a stable fixed point 
at $v_{rest} = -\sqrt{1-I}$ and an unstable ones at $v_u = +\sqrt{1-I}$.
The dynamics is excitable with $v_u$ representing the threshold
to overcome to observe ``an excursion" towards infinity (a spike) before 
relaxing to the rest state at $v_{rest}$ \cite{erme}.
This amounts to saying that if the initial value $v_i(t=0) < v_{rest}$
also at all the successive times the membrane potential will remain 
smaller than $v_{rest}$. While if $ v_{rest} < v_i(t=0) < v_u$
the membrane potential will tend asymptotically to  $ v_{rest}$.
Furthermore, for $ I > 1$ the neuron
fires periodically with frequency $\nu = \sqrt{I-1}/(\pi \tau)$.


Since the network is fully connected, with equal synaptic weights, all neurons
receive the same synaptic current $I(t)$ that is the linear superposition
of all the pulses emitted in the network up to the time $t$.
In particular, as schema for the Post-Synaptic Potentials (PSPs)
we consider step functions of finite duration $T_s$ and amplitude $J \equiv G/(N T_s)$, 
therefore the current reads as: 
\begin{equation} 
\label{PSP}
I(t)= J \sum_{\left\{t_n\right\}} 
\left[ \Theta(t-t_n)-\Theta(T_s+t_n-t)  \right] \quad ,
\end{equation}
where $\Theta(x)$ is the Heaviside function, 
the sum runs over all the spike times $ t_n < t$, and the coupling is normalized
by the number of neurons $N$ 
to ensure that the total synaptic input will remain finite in the limit
 $N \to \infty$. We have consider pulses of the form (\ref{PSP})
as the simplest example of PSPs allowing to take in account spatial and 
temporal summation of stimuli, due to their finite duration and amplitude.

In the limit $T_s \to 0$, the PSPs will become $\delta$-pulses 
and in this case the synaptic current can be rewritten as follows:
\begin{equation} \label{delta}
I(t)= \frac{G}{N}  \sum_{\left\{t_n\right\}} \delta(t-t_n)  
\end{equation}

By following \cite{Hansel_Mato_2001}, we can derive the average
firing rate $\nu$ in infinite size network, in this case the spiking 
frequency of the single neuron is simply given by 
\begin{equation} \label{HM1} 
\nu=\frac{\sqrt{G \nu-1}}{\pi\tau}  
\end{equation} 
where $G \nu$ is the total synaptic current received by each single neuron,
this result is valid both for the step PSPs \eqref{PSP} as well as for the
$\delta$-pulses.
By solving the  implicit equation above one gets \begin{equation}
\label{HM} \nu_{1,2}=\frac{G\pm\sqrt{G^2-4\tau^2\pi^2}}{2\tau^2\pi^2}
\end{equation} 
therefore there are two branches of solutions, we will re-examine
this point later. Let us just mention that 
these solutions have been associated 
with the {\it asynchronous persistent states} emerging in networks 
composed by inhibitory and excitatory QIF populations \cite{Hansel_Mato_2001}.

 A particular example of asynchronous state emerging in globally coupled
networks is the so-called splay state \cite{vvres,Zillmer_et_al_2007}.
This regime is characterized by a sequential firing of all the neurons 
with a constant {\it network interspike interval} (NISI) $T$,
while the dynamics of each neuron is periodic with period  $N \cdot T$.
This state has been classified as an asynchronous regular state \cite{Brunel_2000}.
Splay states have been found in an all-to-all
pulse coupled excitatory network for LIF models~\cite{Zillmer_et_al_2007} 
as well as for general neuronal models \cite{Calamai_et_al_2009}, and in inhibitory
networks for $\delta$-pulses \cite{abbott,Zillmer_et_al_2006}.

\section{Event-driven map}
As previously done in \cite{Jin_2002, Zillmer_et_al_2007} for LIF neuronal models, 
we would like to derive an event-driven map for the setup considered
in the present paper.  The event-driven map gives the exact evolution of the
system, described by the set of $N$ ODEs \eqref{QIF} plus the variable describing
the synaptic current, from an {\it event} to the
successive one. Therefore the continuous time evolution is substituted by a
map with discrete time. 

Let us first consider PSPs that are step pulses of duration $T_s$
as reported in \eqref{PSP}. In the last part of the section we will derive the event-driven map also in the $\delta$-pulses limiting case.

\subsection{Step pulses}
In the case of step pulses, two type of events should be distinguished: pulse emission
(PE) and pulse termination (PT). Both events induce an instantaneous change of
the synaptic current by a constant value: the current will increase (resp. decrease)
by a quantity $J$ for PE (resp. PT). In order to integrate
the system it is not sufficient to know the value of the membrane potentials
and of the synaptic current at a certain time $t$. The system evolution will depend also on the termination
times of the previous pulses received by the neuron that are  ``active" (still contributing to
the synaptic current) at time $t$. Therefore one needs to know the ordered list of 
the future PT times $\{ S_j (t) \}$ with $j=1,\dots, K$, where $ t < S_1(t) <  S_2(t) < \dots < S_K(t)$. 
The number $K(t)$ of these events is in general not constant and it represents the number of 
overlapping pulses at time $t$, which amounts to a  synaptic current $I(t) = K(t)J$. 
Let us now discuss separately how the PE and PT events influence the neural 
dynamics in order to derive an event-driven map. 

\paragraph*{Pulse Emission}

Suppose that at time $t_n$ the neuron $q$ emits a spike and that at time $t_n^-$ 
there were $K$ overlapping pulses. One can obtain the value of
the membrane potential for the neuron $i$ at the next event,
occurring at $t_n + \Delta t$, by integrating equation \eqref{QIF} with $I(t)=(K+1)J$
\begin{equation}
\label{integral_PE}  
\int^{v_i(t_n+\Delta t)}_{v_i(t_n^+)} \frac{d X}{X^2 + (K+1)J-1} 
= \int^{t_n + \Delta t}_{t_n^+} \frac{dt}{\tau}  \qquad.
\end{equation}
How to determine the time interval $\Delta t$ will be explained in the following. Due to the simple form of the PSP  we can
integrate  \eqref{integral_PE} analytically, obtaining
\begin{equation}
\label{F1m} 
v_i(t_n+\Delta t) =\left\{
\begin{array}{l r}
 H(v_i(t_n^+),K+1,\Delta t), & \qquad i \ne q \\ 
&\\
H^*(K+1,\Delta t), & i=q
\end{array}\right. 
\end{equation}
with
\begin{equation}
\label{Hh} 
\begin{array}{l r}
H(x,K,t) =\frac{[KJ-1]\beta_{K}(t)+x}{1-\beta_{K}(t) x }  , &H^*(K,t)=-1/\beta_{K}( t),
\end{array}
\end{equation}
and with the function $\beta_K$ defined as follows
\footnote{Please notice that in the excitable case ($KJ <1$)
one gets a single valued function from the integral
\eqref{integral_PE} due to the fact that, depending on the initial
value of the membrane potential, the dynamics
remains segregated in one of the three intervals
$v_i(t) < v_{rest}$ or $v_{rest} < v_i(t) < v_u$
or $ v_i(t) > v_u$.}
\begin{equation}
\label{beta} 
\beta_K(t)=\left\{ 
\begin{array}{l l l}
K J < 1 && \frac{\tanh \left( \sqrt{1-K J} t/\tau \right)}{\sqrt{1-K J}}
\\
&&\\ 
&&\\ 
K J >  1 && \frac{\tan \left( \sqrt{K J-1} t/\tau \right)}{\sqrt{K J-1}}
\end{array}
\right. 
\end{equation}
Furthermore, the list of the future PT times should be updated by adding
$S_{K+1}(t_n) = t_n + T_s$.

\paragraph*{Pulse Termination}

Let us now consider a PT occurring at time $t_{PT}$ when
there were $K \ge 1$ overlapping pulses present in the network.
The membrane potential of the $i$-th neuron at the next event,
occurring at $t_{PT} + \Delta t$, can be obtained by solving the following integral
\begin{equation}
\label{integral_PT}  
\int^{v_i(t_{PT}+\Delta t)}_{v_i(t_{PT}^+)} \frac{d X}{X^2 + (K-1)J-1} 
= \int^{t_{PT} + \Delta t}_{t_{PT}^+} \frac{dt}{\tau}  \qquad ;
\end{equation}
which gives
\begin{equation}
\label{potential_PT}  
v_i(t_{PT}+\Delta t) = H(v_i(t_{PT}^+),K-1,\Delta t)
\qquad .
\end{equation}
At each pulse termination the list of the PT times $\{ S_j(t_{PT})\}$ should be
updated by throwing away the smallest time $S_1$ and by relabeling the other times as 
$S_{j}(t^+_{PT}) = S_{j+1}(t^-_{PT})$ with $j=1,...,K-1$.

\paragraph*{Determination of the Integration Time-Lapse}

After each event PE or PT at time $t^*$, one should
determine the time interval $\Delta t$ until the next event.
In particular, one should understand if the next event will
be a PE or a PT. In order to resolve this dilemma, the next 
presumed firing time $E(t^*)$ occurring
in the network has to be firstly determined on the basis
of the values of the membrane potentials and of the synaptic current 
at time $t^*$. In the absence of any intermediate event, since
we are considering a fully coupled system, the
neuron $p$ with highest membrane potential value $v_p(t^*)$
is going to fire at time $E(t^*)$. This time can be determined by imposing that  $H(v_p(t^*),K,E(t^*)-t^*)=\infty$, with $H$ given by eq. \eqref{Hh}, namely
\begin{equation}
\label{NISI} 
E(t^*) = \left\{ 
\begin{array}{l l l}
K J <1 && t^*+ \frac{\tau}{\sqrt{1-K J}} \left[ \tanh^{-1}\left(\frac{\sqrt{1-K J}}{v_p(t_n^+)}\right)\right]\\
&&\\
&&\\ 
K J  >1 && t^*+ \frac{\tau}{\sqrt{K J-1}} \left[ \tan^{-1}\left(\frac{\sqrt{ KJ-1}}{v_p(t_n^+)}\right)\right]\\
\end{array}\right. 
\end{equation}
where $K$ is the number of overlapping pulses immediately after the event at $t^*$.
In order to understand the type of the next event
$E(t^*)$ should be compared with $S_1(t^*)$
to determine which is the smaller one. If $K=0$ then $\Delta t=E(t^*)-t^*$ automatically, otherwise

\begin{equation}
\Delta t=\min\left\{E(t^*),S_1(t^*)\right\}-t^* \quad.
\end{equation}
The event-driven map will be therefore a combination of the two above 
described integration steps. 
After each event
the potential will be given by eq. \eqref{F1m} or 
eq. \eqref{potential_PT} depending if the event is a PE or a PT.

\paragraph*{Co-moving frame}

A further simplification to the above scheme can be obtained by 
exploiting the fact that for globally coupled networks
the neuron firing order is preserved. Since the firing order
is directly related to the membrane potential value, 
we can order sequentially the membrane potentials, i.e. 
$v_1(t) > v_2(t) > \dots > v_N(t)$, and introduce
a co-moving frame. This amounts to relabeling the neuron 
closest-to-threshold as $1$, and when it fires
at time $t_n$ to reset the potential value as $v_1(t_n^-) \to v_N(t_n^+)=-\infty$
and to shift the indexes of all the others $i \to (i-1)$ for $i \ge 2$.
Furthermore, due to the reference frame transformation, 
Eq. \eqref{F1m} has to be modified: namely
the evolution map should be rewritten as
$v_{i}(t_n+\Delta t) = H(v_{i+1}(t_n^+),K+1,\Delta t)$, for $i=1, \dots, N-1$
and $v_N(t_n+\Delta t)=H^*(K+1,\Delta t)$.

\subsubsection{Splay-state}
\begin{figure}[htbp]
\begin{center}
\includegraphics[draft=false,clip=true,height=0.75\textwidth]{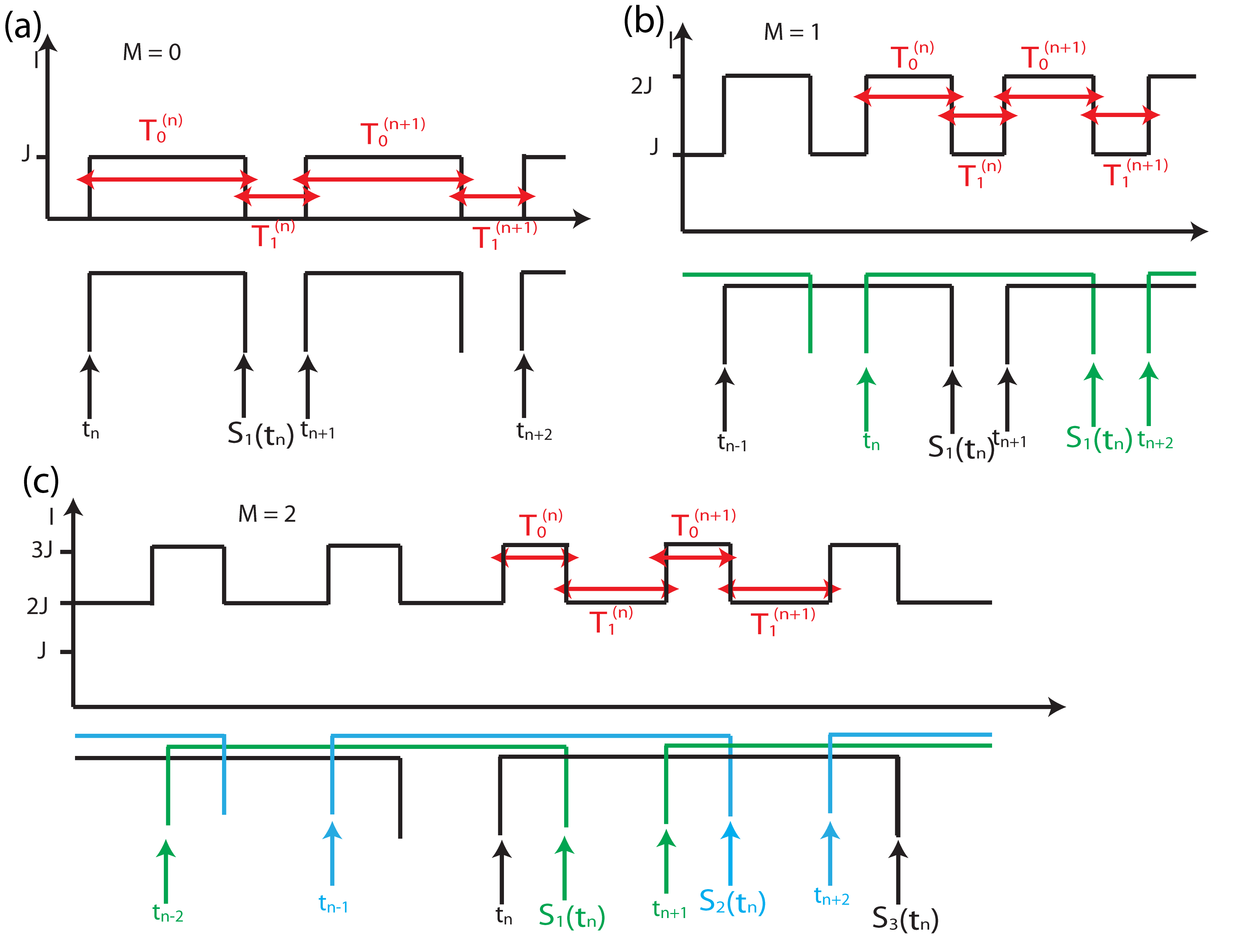}
\end{center}
\caption{PSPs in a splay state can overlap $M$ times. (a) PSPs overlapping $M=0$ times (no overlaps);
(b) PSPs overlapping $M=1$ times;  (c) PSPs overlapping $M=2$ times. Independently 
of the value of $M$ the synaptic current can take only two values in the
time interval between two spikes: namely,
during $I(t_n<t<t_n+T_0^{(n)})=(M+1)J$  and $I(t_n+T_0^{(n)}<t<t_{n+1})=M J$. 
}
\label{fig1}
\end{figure}

For the splay state regime, the event-driven map outlined above simplifies
noticeably and furthermore it can be explicitly written.
The splay state is characterized by a constant NISI: $T$. 
Furthermore, due to the regular spike emission the PT times can be
all written in function of $S_1(t)$ as $ S_j(t)= S_1(t) + (j-1) \cdot T$.
In general, it is useful to rewrite $T_s$ as a function of $T$, as follows
\begin{equation} 
\label{t0m} 
T_s=M T+ T_0 
\end{equation}
where $K=M$ is the number of overlapping PSPs just before the spike emission,
$T_0<T$ and let us define $ T_1=T- T_0$. Please notice that
for a splay state $K$ can assume only two values, namely $M$ and $M+1$
as shown in Fig. \ref{fig1}.
In the case of non-overlapping pulses: $M=0$, $T_s \equiv  T_0$, and
$T_1 \equiv T-T_s$. This case is illustrated in Fig. \ref{fig1}(a).

In order to determine the value of the coupling $G_M$ required
to have exactly $M$ overlapping pulses, 
let us employ, as a first approximation, the mean field equation \eqref{HM} with the condition 
$\nu_1=M/(NT_s)$, which is equivalent to assuming that $T_s \equiv M T$, 
\begin{equation} 
\label{HMm} 
\frac{M}{NT_s}=\frac{G_M+\sqrt{G_M^2-4\tau^2\pi^2}}{2\tau^2\pi^2} 
\quad ,
\end{equation}
then we can invert the above equation and obtain the critical coupling
\begin{equation} 
\label{gm} 
J_M = \frac{G_M}{N T_s} = \frac{1}{M} + \frac{\tau^2\pi^2 M}{(N T_s)^2}
\quad .
\end{equation}
If $J < J_1$ there is no overlap among two successive emitted PSPs.
When $J_M< J <J_{M+1}$, $M$ pulses overlap.
The synaptic current can take only the following two values
\begin{equation}
\label{Isyn_splay} 
I(t) = \left\{ 
\begin{array}{l l l}
(M+1) J && \qquad t_n < t < t_n + T_0
\\
&&\\ 
M J && \qquad t_n + T_0 < t < t_{n+1}
\\
\end{array}\right. 
\qquad ;
\end{equation}
as clearly illustrated in Fig. \ref{fig1}. In particular, if $T_0=0$,
we will have always exactly $M$ overlapping pulses, since each PE
will coincide with a PT, and $I = M J$.

For the splay state we can rewrite the dynamics of each neuron $i$ 
between two successive spikes occurring at $t_n$ and $t_{n+1}$ as an 
exact map made of the following three steps.

\begin{enumerate}

\item The first step starts with a PE at time $t_n$,
one can easily estimate the evolution of the membrane potential
from time $t_n^+$ to $T_1$ when a PT will occur. Let us
first define $x_i^{(n)} = v_i(t_n^+)$ and $y_i^{(n)} = v_i(t_n^++T_0)$,
and order the membrane potentials as follows
\begin{equation}
\label{conditionX} x^{(n)}_1>x^{(n)}_2>...>x^{(n)}_{N}=-\infty
\qquad ,
\end{equation}
the last equivalence stems from the fact that a neuron has just fired
and it has been reset.
By employing the expression \eqref{F1m} one gets the following map
\begin{equation}
\label{map1} 
y_i^{(n)}=\left\{
\begin{array}{l r}
  F_1 ( x_i^{(n)}, T_0 ) =H(x^{(n)}_i,M+1,T_0), & i \neq N\\
&\\ 
F^*_1(T_0 ) = H^*(M+1,T_0),& i=N
\end{array}\right. 
\end{equation}
with $H$ and $H^*$ defined in \eqref{Hh}.

\item The second step corresponds to the integration of the equation
of motion from the PT occurring at $t_n+T_0$ and the time $t_{n+1}^-$
immediately preceding the $n+1$-th spike emission.
By defining $z_i^{(n)} = v_i(t_{n+1}^-)$ and by employing equation
\eqref{potential_PT} one gets
\begin{equation}
\label{map2} 
z^{(n)}_i =H(y^{(n)}_i,M,T_1)
\end{equation}
with $H$ defined in \eqref{Hh}. Due to the previous ordering,
the next firing neuron will have the label $1$, therefore
$z^{(n)}_1=\infty$ and thus the denominator of the right side equation
\eqref{map2} should be zero:
\begin{equation}
\label{condition} 
1-\beta_M(T_1) y_1^{(n)}=0 
\end{equation}
By inserting \eqref{condition} in \eqref{map2} one gets:
\begin{equation}
\label{F2m_bis} 
z^{(n)}_i=F_2(y^{(n)}_1,y^{(n)}_i) =
\frac{(MJ-1) +y^{(n)}_1 y^{(n)}_i}{y^{(n)}_1-y^{(n)}_i} 
\end{equation} 

\item The last step amounts simply to calculating the membrane potential change in
going from $t_{n+1}^-$ to $t_{n+1}^+$ and introducing a co-moving frame to maintain 
the order among the membrane potentials also after each firing event.
This amounts to writing
\begin{equation}
\label{F3} x^{(n+1)}_{i}=F_3(z^{(n)}_{i+1})=z^{(n)}_{i+1}  \qquad {\rm for} \enskip 1 \le i \le N-1
\end{equation}
and setting $x_N^{(n+1)} = -\infty$.
Since the event-driven map approach corresponds to a suitable Poincar\'e section,
we are left with $N-1$ variables, dropping the variable $i=N$.

\end{enumerate}

We can compute the complete event-driven map from spike time $t_n$ to spike time $t_{n+1}$, 
by combining the three above equations \eqref{map1}, \eqref{map2} and \eqref{F3} 
\begin{equation}\label{Fm}  
x^{(n+1)}_{i}=F(x^{(n)}_{i+1})=\frac{a_0+a_1 x^{(n)}_{i+1}}{a_2+a_3 x^{(n)}_{i+1}}  \qquad {\rm for} \enskip 1 \le i \le N-1
\quad ;
\end{equation}
where the coefficients entering in \eqref{Fm} reads as:
\begin{equation}
\label{coef1m} 
\begin{array}{l} 
a_0=(MJ-1)\beta_M(T_1)+[(M+1)J-1]\beta_{M+1}(T_0)\\
a_1=1-(MJ-1)\beta_{M+1}(T_0)\beta_M(T_1) \\ a_2=1- [(M
+1)J-1]\beta_{M+1}(T_0)\beta_M(T_1)  \\ 
 a_3=-\beta_{M+1}(T_0)-\beta_M(T_1)
\end{array}  
\end{equation}

\paragraph*{Exact firing rate value.} In order to obtain the membrane potential values associated with 
the splay state one should impose that the splay state represents a fixed point 
for the event-driven map in the comoving frame, namely
\begin{equation}
\label{fix_point} 
x_i^{(n)} = x_i^{(n+1)} = \tilde x_i
\qquad 
\end{equation}
Furthermore, once fixed $G$, $N$, and $T_s$, one
can determine the NISI $T$ by solving iteratively
equation  \eqref{NISI} together with the set of equations
for the membrane potential \eqref{Fm}, with the requirement
that $x^* = F^N(\tilde x_N=-\infty) = +\infty$. 
Numerically, as a first guess for $T$ we usually employ the mean-field
result $1/\nu_1$, given by the larger solution of \eqref{HM}. Then we evaluate the splay state by 
employing a bisection method to find the exact NISI. We stop the procedure whenever $x^* > 10^8$, with the constraint that 
the order \eqref{conditionX} is maintained.

\begin{figure}[htbp] \begin{center}
\includegraphics[draft=false,clip=true,height=0.75\textwidth]{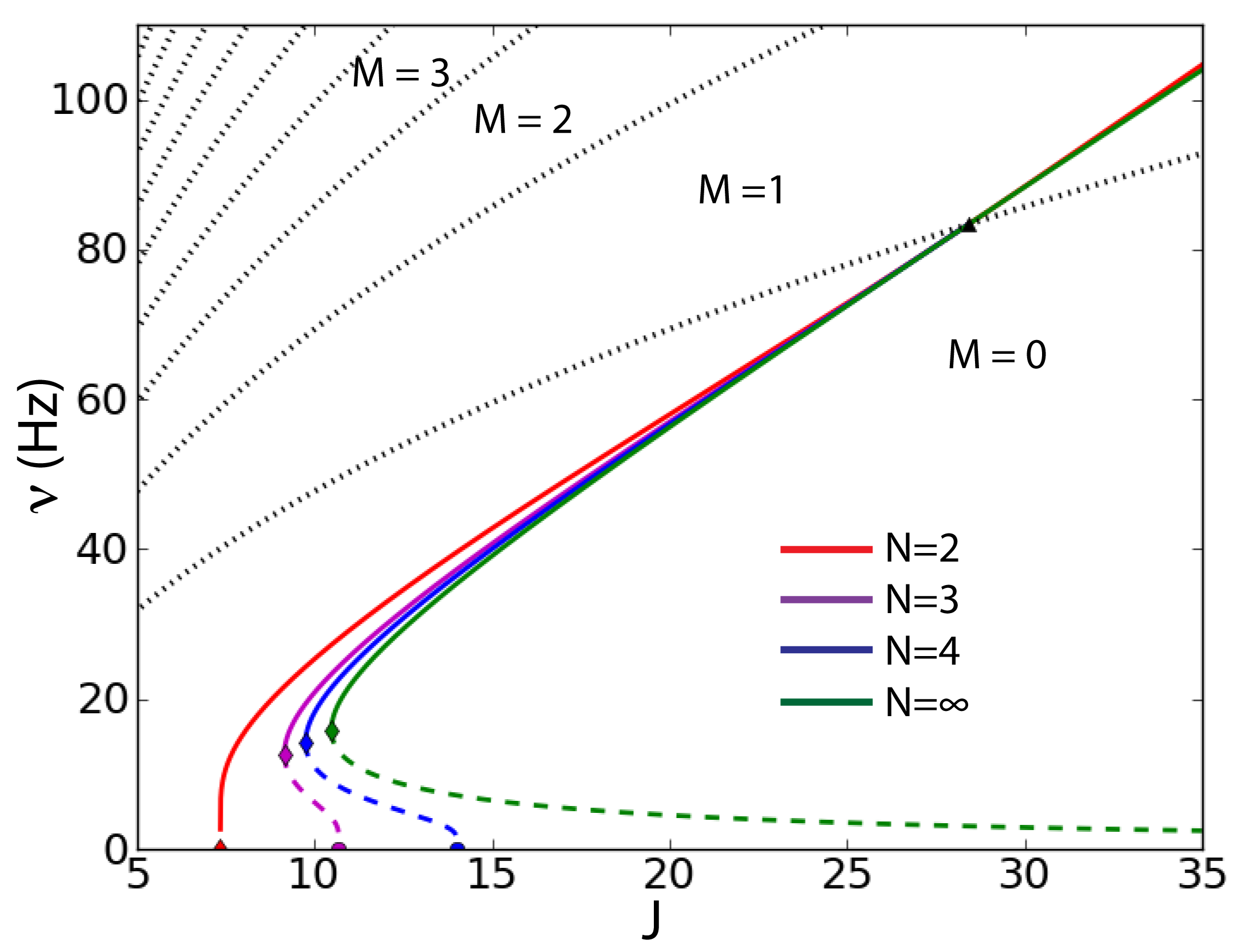}
\end{center}
\caption{Frequencies of the splay states as a function of the synaptic
strength $J$ and for pulse duration $T_s N=12$ ms with $\tau=20$ ms.
Red line: $N=2$, magenta line: $N=3$, blue line: $N=4$, green line:
$N=\infty$. Black dotted lines separate regions with different number $M$ of overlapping PSPs. 
Solid lines refer to the upper stable branches of the splay state.
Dashed lines refers to the lower unstable branches of the splay state. For $N=2$ the lower branch does
not exist.} 
\label{fig:fvsg} 
 \end{figure}

For a given set of parameters $G$, $T_s$ and $N$ we found at maximum
two coexisting splay states (in agreement with the mean-field results). 
Beyond a minimal value of $J$, there is always one marginally stable splay state. When the splay states are two we found that the other one is
 unstable, as illustrated in the following in Fig. \ref{fig:FiguraK}.
Let us stress that unstable branches of solutions exist only for
non overlapping pulses (i.e. $M=0$) as shown in Fig.  \ref{fig:fvsg}.
These numerical results will be
confirmed by analytical analysis in \S A for $N=2,3,4$ and $J<J_1$ ($M=0$).

Notice that for $N=2$ only the marginally stable branch exists
and the minimal firing rate reaches the value $\nu=0$. Instead for $N>2$
the minimal firing rate of the marginally stable branch is $\nu\neq 0$.
The firing rates associated to the unstable branch, for finite $N$, 
reaches always the value $\nu=0$ for some finite pulse amplitude $J=J^*$. 
Finally, for $N\to\infty$ we have that $J^*(\nu=0)\rightarrow\infty$.

\subsection{$\delta$-pulses}
In the case of $\delta$-pulses, the formulation of the event-driven map is
extremely simplified, since now there are only PE events. 
At the arrival of a $\delta$-pulse, we can integrate eq. \eqref{QIF} 
with the current given by \eqref{delta} between time $t^-_{n}$ and $t^+_n$, obtaining 
\begin{equation}
\label{map2_d} 
y^{(n)}_{i} = x^{(n)}_i + J_{\delta} \qquad {\rm for} \enskip 1 \le i \le N-1 \quad ,
\end{equation} 
where $J_{\delta}=G/(N\tau)$.
The evolution of the membrane
potential in the time interval $t_n^+$ and $t_{n+1}^-$ can be easily obtained 
since it corresponds to eq. \eqref{potential_PT} with $M=0$ and $T_1=t_{n+1}^--t_n^+=T$, namely
\begin{equation}
\label{integral_d1}  
z_i^{(n)}= H(y^{(n)}_i,0,T)
\end{equation}
for $i=1,\dots,N-1$.  Then we can combine eq. \eqref{map2_d} and \eqref{integral_d1} with the change of reference frame \eqref{F3} to obtain the corresponding event-driven map.
The resulting map is identical to that found for the step function
\eqref{Fm}, apart the value of the coefficients \eqref{coef1m} that now
become:
\begin{equation}
\label{coef22} \begin{array}{l}  
a_0=-\beta_0(T) + J_{\delta} \\ 
a_1=1 \\ 
a_2 =1-\beta_0(T) J_{\delta} \\  
a_3=-\beta_0(T) 
\end{array} 
\end{equation}
Once fixed $J_{\delta}$ and $N$ and $T_s$, 
similarly to the case of step pulses, one can determine $T$ 
together with the membrane potential values associated with the splay state
by solving iteratively equation \eqref{NISI}, 
and by applying iteratively the map \eqref{Fm} with coefficients \eqref{coef22} 
starting from $\tilde x_N=-\infty$. The solution is numerically achieved 
whenever $x^* = F^N(\tilde x_N=-\infty) = +\infty$ (namely, $x^* > 10^8$) and
the condition \eqref{conditionX} is satisfied.

We want to conclude this Section by mentioning the fact that in the
limit $N \to \infty$ we were able to derive an explicit analytic expression
for the membrane potentials corresponding to a splay state. The detailed
calculations are reported in \S B.

\section{Linear Stability Analysis for Step Pulses}
We are interested in the linear stability of the splay state 
in the case of step pulses for finite system size $N$. 
It is therefore useful to introduce the following vector notation for the membrane potentials at spike time $t_n$:
\begin{equation}
\label{xvector} 
\mathbf{x}^{(n)}= \left\{ x^{(n)}_1, x^{(n)}_2, \cdots, x^{(n)}_{N}\right\} \qquad ;
\end{equation}
Furthermore, if we have more than one overlapping pulse, i.e. if $M>0$, the actual
state of the network will depend not only on the membrane potential
values but also on the past $M$ spike times $\{t_k\}$ with
$k=n-M, n-M+1, \dots,n-1$. However the formulation of the tangent space
dynamics can be made simpler by introducing the 
related time intervals $\tau^{(n)}_j \equiv t_n - t_{n-j}$:
\begin{equation}
\label{xvector1} 
\mathbf{\tau}^{(n)}= \left\{\tau^{(n)}_1,
\tau^{(n)}_2, \cdots, \tau^{(n)}_{M}\right\} 
\end{equation}

In this notation the splay state is a fixed point of the network dynamics satisfying the 
following relationships:
\begin{equation}
\label{fix2} 
\tilde{\mathbf{x}}=F_3(\tilde{\mathbf{z}})=F_3(F_2(\tilde{\mathbf{y}}))
=F_3(F_2(F_1(\tilde{\mathbf{x}})))
\quad ,
\end{equation}
and
\begin{equation}
\label{fix1} 
{\tilde \tau}_j = j \cdot T \qquad j=1,\dots,M
\qquad .
\end{equation}


\subsection{Linearized Poincar\'{e} map}
In order to derive the equations of evolution in the tangent space for our case
it is convenient to consider separately the three steps in eq. \eqref{fix2}, 
please notice that now $T_0^{(n)}$ and $T_1^{(n)}$ depend on the spike sequence index $n$
since, for the perturbed dynamics
these quantities are no more constant.

Let us start by perturbing eq. \eqref{map1}:
\begin{equation}
\label{perturb1}
\left\{\begin{array}{l l} 
\delta y^{(n)}_{i=1,...,N-1}&=d_i\delta x_i^{(n)}+s_i\delta T^{(n)}_0\\ 
&\\ 
\delta y^{(n)}_N&=s_N\delta T^{(n)}_0 
\end{array}
\right.
\end{equation}
with $\delta T^{(n)}_0=0$ if $M=0$;
where the coefficients are:

\begin{equation}
\label{coef_d} 
d_i=\left.\frac{\partial F_1(x^{(n)}_i,T^{(n)}_0)}{\partial x^{(n)}_i}\right|_{\tilde{x}_i,
\tilde{T}_0}=\frac{1+[(M+1)J-1]\beta^2_{M+1}(\tilde{T}_0)}
{(1-\beta_{M+1}(\tilde{T}_0) \tilde{x}_i)^2}  
\qquad ,
\end{equation}

\begin{equation}
\label{coef_s}
s_i=\left.\frac{\partial F_1(x^{(n)}_i,T^{(n)}_0)}{\partial T^{(n)}_0}\right|_{\tilde{x}_i,
\tilde{T}_0}=\frac{(M+1)J-1+\tilde{x}^2_i}{(1-\beta_{M+1}(\tilde{T}_0)\tilde{x}_i)^2}
\frac{1}{\cos^2\left(\sqrt{(m+1)g-1}\tilde{T}_0/\tau\right)}\frac{1}{\tau}
\qquad ,
\end{equation}

\begin{equation}
\label{coef_sN}
s_N=\left.\frac{d F^*_1(T^{(n)}_0)}{d T^{(n)}_0}\right|_{\tilde{T}_0}=
\frac{1}{\beta^2_{M+1}(\tilde{T}_0)}\frac{1}{\cos^2\left(\sqrt{(M+1)J-1}\tilde{T}_0/\tau\right)}
\frac{1}{\tau} \qquad ;
\end{equation}

As a second step we perturb $F_2$ given by eq. \eqref{F2m_bis}, obtaining
\begin{equation}
\label{perturb2}
\delta z^{(n)}_i=h_i \delta y^{(n)}_1+ k_i \delta y^{(n)}_i 
\qquad i=1,\dots,N \quad ;
\end{equation}
remember that if $M=0$ then  $\delta y^{(n)}_N=0$. The coefficient $h_i$ and
$k_i$ are defined as:
\begin{equation}
\label{coef_h} 
h_i=\left.\frac{\partial
F_2(y^{(n)}_1,y^{(n)}_i)}{\partial
y^{(n)}_1}\right|_{\tilde{y}_1,\tilde{y}_i}=-\frac{MJ-1+\tilde{y}^2_i}{(\tilde{y}_1-\tilde{y}_i)^2}
\qquad ,
\end{equation}

\begin{equation}
\label{coef_k} 
k_i=\left.\frac{\partial
F_2(y^{(n)}_1,y^{(n)}_i)}{\partial
y^{(n)}_i}\right|_{\tilde{y}_1,\tilde{y}_i}=
\frac{MJ-1+\tilde{y}^2_1}{(\tilde{y}_1-\tilde{y}_i)^2}
\qquad .
\end{equation}

Finally, the linearized equations associated with the 
reference frame change can be obtained by
perturbing eq. \eqref{F3}:
\begin{equation}
\label{perturb3} 
\delta x_{i}^{(n+1)}=\delta z_{i+1}^{(n)}
\qquad  i=1, \dots, N-1 \qquad,
\end{equation}
please notice that $\delta x_{N}^{(n)} \equiv 0$ due to the fact that 
in the comoving frame $x_{N}^{(n)} \equiv -\infty$, therefore the evolution
in the tangent space should deal with only 
$N-1$ perturbations associated to the membrane potentials.

Then we need to compute how the time interval $T^{(n)}_0$ is modified by the 
perturbations, when $M>0$.
The key point here is that $T^{(n)}_0$ depends on the previous spike times as follows:
\begin{equation}
T^{(n)}_0 = T_s -(t_n - t_{n-M}) = T_s - \tau_M^{(n)} \qquad ;
\label{rel1} 
\end{equation}
apparently one could be lead to think that we need only an 
extra variable: $\tau^{(n)}_M$. However, $\tau^{(n)}_M$
depends on all the $M$ previous spike times and therefore
we need to take in account also the perturbations of the
other $M-1$ variables, namely  $\tau^{(n)}_{j=1,...,M-1}$.

To obtain the evolution equations for these auxiliary $M$ variables,
let us consider the following relations
\begin{equation}
\label{tau1}
\tau^{(n+1)}_1=T^{(n)}_0+T^{(n)}_1
\end{equation}
and
\begin{equation}
\label{taui}
\tau^{(n+1)}_j=\tau^{(n+1)}_1+\tau^{(n)}_{j-1}=T^{(n)}_0+T^{(n)}_1+\tau^{(n)}_{j-1}
\quad .
\end{equation}

From \eqref{rel1} we obtain
the relation $\delta \tau^{(n)}_M=-\delta T^{(n)}_0$. 
From this relation and from equations \eqref{tau1} and \eqref{taui} (for positive $M$) we can
easily obtain the evolution maps for the perturbed quantities:
\begin{equation}
\left\{\begin{array}{l l}
\label{dtau1}
\delta \tau^{(n+1)}_1&=\delta T^{(n)}_1-\delta \tau^{(n)}_M\\ &\\ \delta
\tau^{(n+1)}_{j=2,...,M}&=\delta T^{(n)}_1+\delta \tau^{(n)}_{j-1}-\delta
\tau^{(n)}_M 
\end{array}
\right.
\end{equation}

We are left just with the determination of $\delta T^{(n)}_1$, this can be
derived by remembering that the time from the last PT
until the next PE can be calculated by employing eq. \eqref{NISI} with $K=M$, $v_p(t^+-n)$, and $E(t^*)-t^*=T^{(n)}_1$:
\begin{equation} 
T^{(n)}_1=G(y^{(n)}_1)
=\left\{
\begin{array}{l l}
MJ<1 & \frac{\tau}{\sqrt{MJ-1}}\tanh^{-1}\left(\frac{\sqrt{MJ-1}}{y^{(n)}_1}\right)\\
\\
MJ>1 & \frac{\tau}{\sqrt{MJ-1}}\tan^{-1}\left(\frac{\sqrt{MJ-1}}{y^{(n)}_1}\right)\\
\end{array}
\right.
\end{equation}
and
\begin{equation} 
w=\left.\frac{d G}{d y^{(n)}_1}\right|_{\tilde{y}_1}
=-\frac{\tau}{\tilde{y}^2_1+|MJ-1|}
\qquad ;
\end{equation}
and we can obtain:
\begin{equation}
\label{Tdw}
\delta T^{(n)}_1=w \delta y^{(n)}_1 \qquad.
\end{equation}

By combining eq. \eqref{perturb1}, \eqref{perturb2}, \eqref{perturb3}, \eqref{dtau1}, and \eqref{Tdw}, the complete map evolution in the tangent space can be finally
written as follows:
\begin{equation}
\left\{\begin{array}{l l}
\label{fin2} 
\delta x_{i=1,...,N-2}^{(n+1)}&=p_{i+1} \delta x_1^{(n)}+ q_{i+1}  \delta
x_{i+1}^{(n)}+u_{i+1}\delta \tau^{(n)}_M\\ 
&\\ 
\delta x_{N-1}^{(n+1)}&=p_{N}
\delta x_1^{(n)}+u_N\delta \tau^{(n)}_M\\ &\\ \delta \tau^{(n+1)}_1&=r_1\delta
x_1^{(n)}+r_2\delta \tau^{(n)}_M\\ &\\ \delta
\tau^{(n+1)}_{j=2,...,M}&=r_1\delta x_1^{(n)}+\delta \tau^{(n)}_{j-1}+r_2\delta
\tau^{(n)}_M 
\end{array}
\right.
\end{equation}
where we have set $p_i=h_id_1$, $q_i=k_id_i$,  $u_i=-(h_i s_1+k_i s_i)$, $r_1=w
d_1$, and $r_2=-(1+w s_1)$.

In order to determine the stability of the splay state we should compute the
Floquet spectrum by setting
\begin{equation}
\label{final}
\left( 
\begin{array}{c} \delta x^{(n+1)}_1 \\ ...\\ \delta x^{(n+1)}_{N-1} \\
\delta \tau^{(n+1)}_1\\ \vdots\\ \delta \tau^{(n+1)}_M \end{array}
\right)=\mu_l \left( \begin{array}{c} \delta x^{(n)}_1 \\ ...\\ \delta
x^{(n)}_{N-1} \\ \delta \tau^{(n)}_1\\ \vdots\\ \delta \tau^{(n)}_M 
\end{array}
\right)
\end{equation}
where $\mu_l={\rm e}^{\lambda_l +i \omega_l}$ ($l=1,\dots,N+M-1$)
are the so called (complex) Floquet multipliers, while $\lambda_l$ 
(resp. $\omega_l$) are real numbers 
termed Floquet exponents (resp. frequencies). If $||\mu_l|| < 1$ $\forall l$ 
(resp. $||\mu_k|| > 1$ for at least one $k$) the splay state
is stable (resp. unstable). Whenever the largest modulus of the Floquet
multipliers is exactly one the system is marginally stable.

The Floquet spectrum can be obtained by solving the following characteristic polynomial, obtained from eq. \eqref{fin2}:
\begin{equation}
\label{polybig}
\begin{array}{c}
\left(\mu^{N-1}_l-p_2\mu^{N-2}_l-\sum^N_{k=3}p_k\left(\prod^{k-1}_{j=2}q_j\right)\mu^{N-k}_l\right)\left(\mu^M_l-r_2\sum^{M-1}_{k=0}\mu^k_l\right)+\\
\\
+\left(u_2\mu^{N-2}_l+\sum^N_{k=3}u_k\left(\prod^{k-1}_{j=2}q_j\right)\mu^{N-k}_l\right)\left(-r_1\sum^{M-1}_{k=0}\mu^k_l\right)=0
\qquad ,
\end{array}
\end{equation}
which admits $N+M-1$ solutions.

\begin{figure}[htbp] 
\begin{center}
\includegraphics[draft=false,clip=true,height=0.3\textwidth]{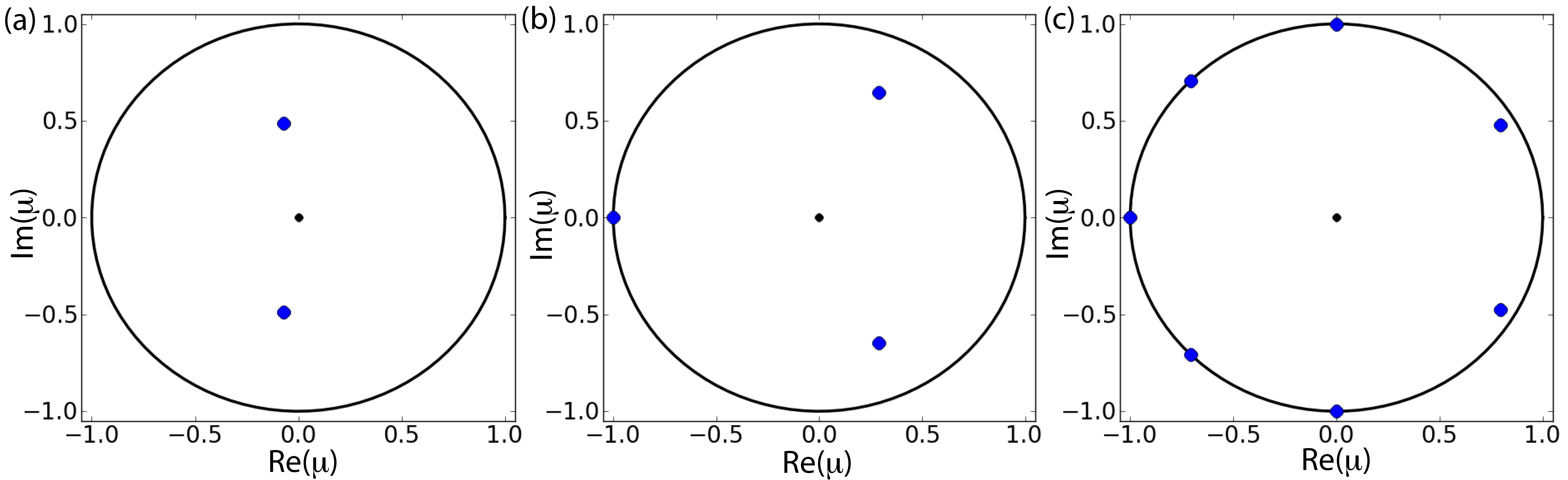}
\end{center} 
\caption{Floquet multipliers $\{ \mu_l \}$ for the case with no overlap,
i.e $M=0$: (a) $N=3$, $0$ marginally stable eigenvalue; (b) $N=4$, $1$
marginally stable eigenvalue; (c) $N=8$, $5$ marginally stable eigenvalues.  
In this case we fixed $J=15$ and $N T_s =16$ ms and we vary the network
size.}
\label{fig:Nincr} 
\end{figure}


\subsection{Floquet multipliers}
\begin{figure}[htbp] 
\begin{center}
\includegraphics[draft=false,clip=true,height=0.3\textwidth]{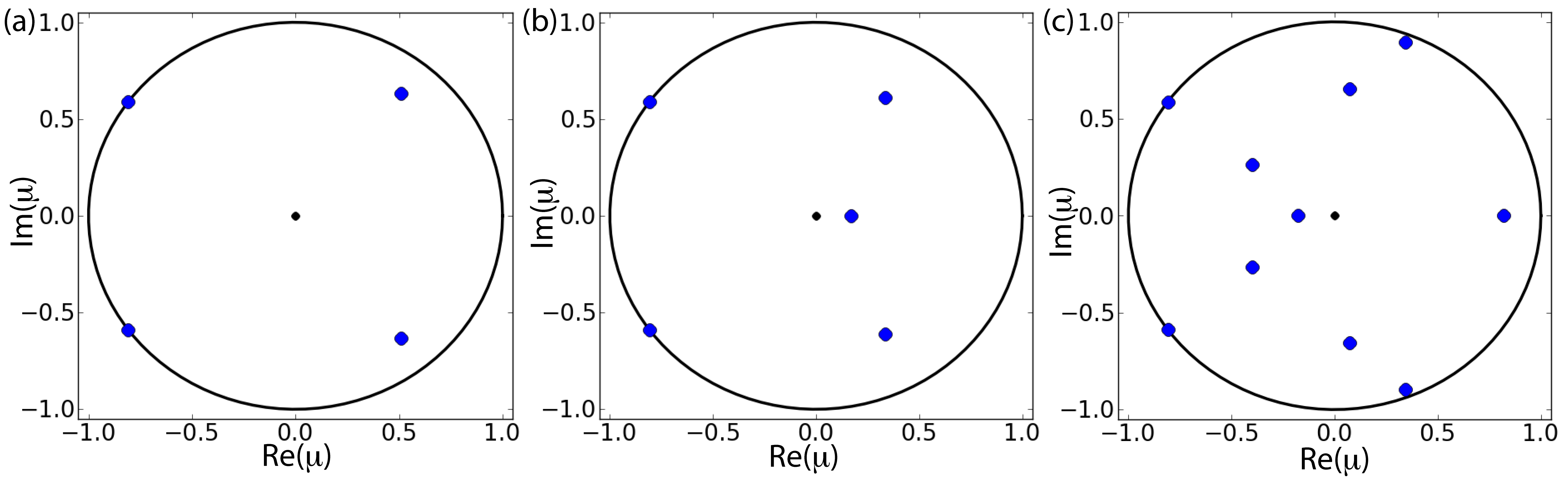}
\end{center}
\caption{
Floquet multipliers $\{ \mu_l \}$ for overlapping pulses, i.e. $M >0$ :
(a) $J=15$, $M=0$, $2$ neutrally stable eigenvalues; (b) $J=25$, $M=1$,
$2$ neutrally stable eigenvalues; (c) $J=100$, $M=6$,  $2$ neutrally stable
eigenvalues.  We have considered $N=5$ and $T_s =3.2$ ms.
} 
\label{fig:Mincr} 
\end{figure}

\begin{figure}[htbp] \begin{center}
\includegraphics[draft=false,clip=true,height=0.32\textwidth]{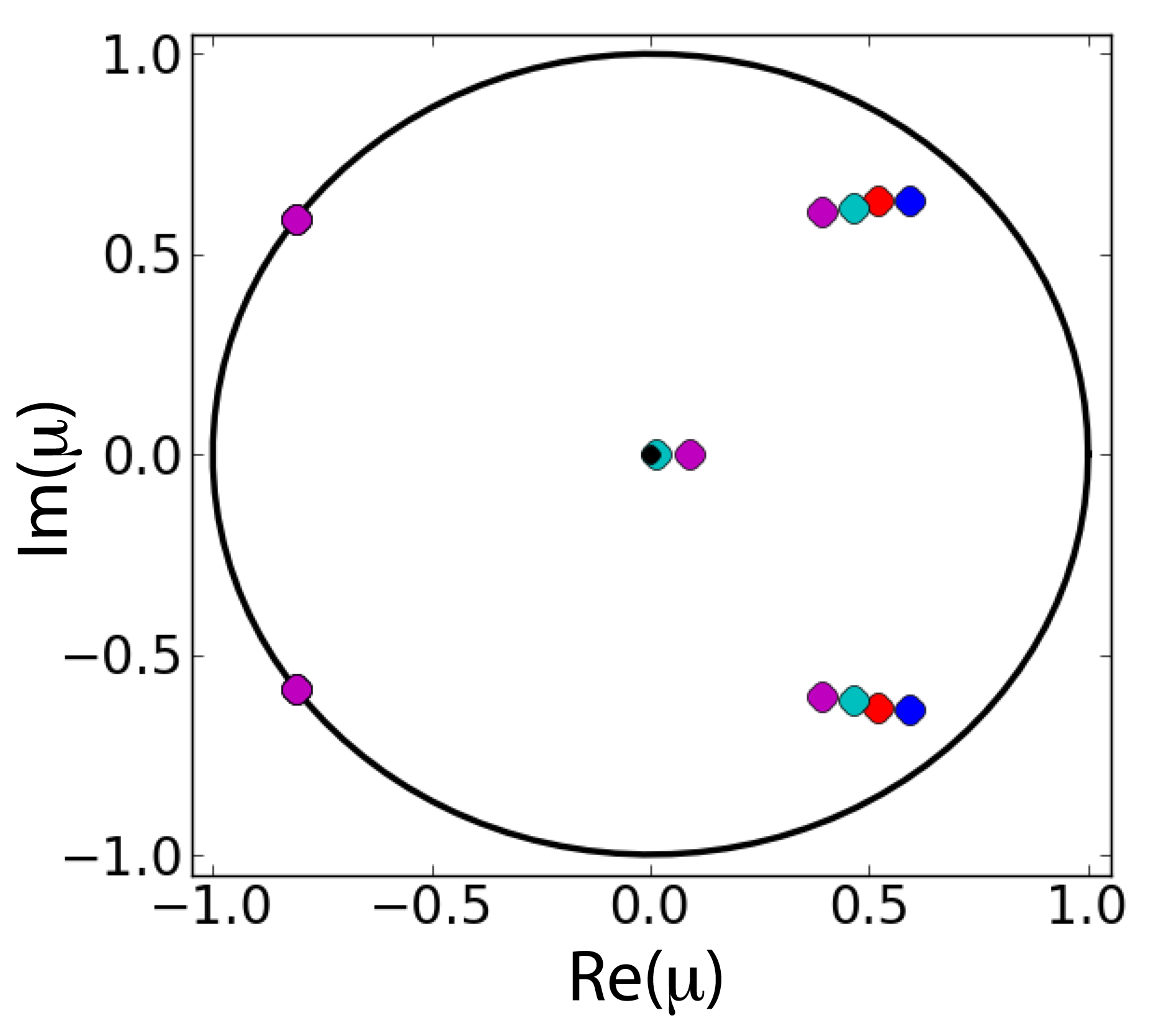}
\end{center}
\caption{(a) Floquet spectrum of the splay state in the complex plane 
for  $T_s N=16$ ms, $N=5$, in this case $J_1=16.42$. Blue stars correspond to 
$M=0$ when $J=10.42<J_1$ (blue), and $J=14.42<J_1$ (red); and to $M=1$ when $J=18.42>J_1$ (cyan), and $J=22.42>J_1$ (magenta)} 
\label{fig:EigenBorn} 

\end{figure}

As stated by Watanabe and Strogatz (WS) \cite{watanabe} 
for a network on $N$ fully coupled phase oscillators with sinusoidal
coupling, the system has in general $N-3$ marginally stable directions,
furthermore for a splay state, which is a periodic solution,
these directions reduce to $N-2$.
Therefore since also our model, as detailed in \S C,
satisfies the hypothesis for which the WS results apply,
and since in the event-driven map formulation one
degree of freedom is lost,
we expect that for the splay states
at least $N-3$ Floquet multipliers lie on the unit circle,
as shown in Fig. \ref{fig:Nincr} for $M=0$. Furthermore, 
in presence of overlaps, i.e. for $M >0$, 
the Floquet exponents associated to the auxiliary variables
${\bf \tau}^{(n)}$ do not influence the stability of the
splay state, since these additional $M$ exponents are
located within the unit circle, and therefore associated
to stable directions as shown in Fig. \ref{fig:Mincr} 
and \ref{fig:EigenBorn}.

It is interesting to notice how the additional exponents associated
to the auxiliary variables emerge by increasing the number of overlaps.
In particular, the number of overlaps can be increased from $M$ to $M+1$
by varying the coupling $J$ from below to above the threshold $J_{M+1}$. At the threshold $J_{M+1}$ a new 
variable $\tau_{M+1}$ is added to the event-driven map describing 
the system. Therefore the Floquet spectrum associated with the corresponding splay state
solution has one additional eigenvalue. This new direction emerges
as superstable at $J = J_{M+1}$ being associated to a zero Floquet multiplier,
as shown in Fig. \ref{fig:EigenBorn}. By further increasing $J$ the new
eigenvalue increases its modulus, which however remains always smaller than
one.

\begin{figure}[htbp] \begin{center}
\includegraphics[draft=false,clip=true,height=0.34\textwidth]{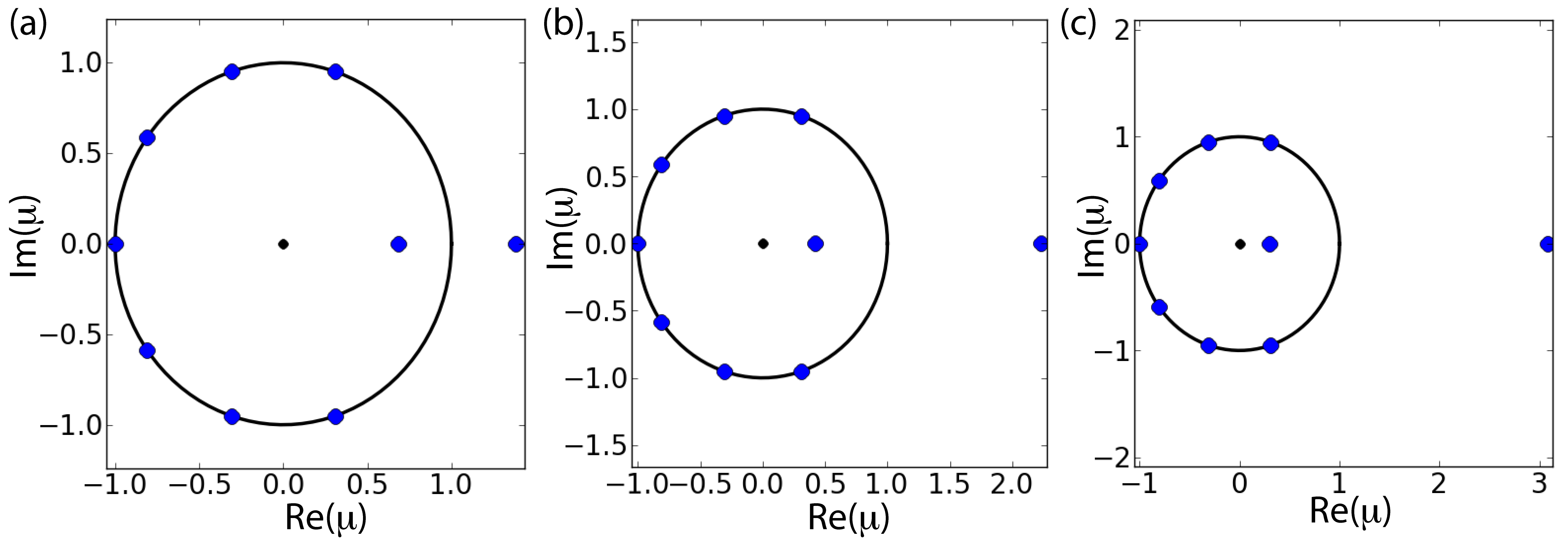}
\caption{Floquet spectrum of the splay state in the complex plane for the unstable branch,
 $T_s N=16$ ms, $N=10$: (a) $J=8$, (b) $J=10$, (c) $J=12$.} 
\label{fig:FiguraK} 
\end{center}
\end{figure}

In Fig. \ref{fig:FiguraK} we report the Floquet multipliers associated to the
unstable branch of splay state solutions, which coexist with the marginally 
stable branch for $N>2$, as already mentioned in Sect. III A.

\section{Linear Stability for $\delta$-pulses}
In the case of $\delta$-pulses the stability of the splay state can be
inferred by theoretical arguments based on the symmetry of the
considered model and of the specific pulse coupling.
It is evident that the QIF model \eqref{QIF} for time symmetric pulses
has a time reversal symmetry. This can be appreciated as follows.
Given a solution $\mathbf{v}(t)=\{v_1(t),\ldots , v_N(t)\}$ we define
$\mathbf{w}(t)=\{w_1(t),\ldots,w_N(t)\}= -\{v_N(-t),\ldots , v_1(-t)\}$. It is clear from 
the time reversal property of
\eqref{QIF} that $\mathbf{w}(t)$ is a solution in between two spike emissions.
Let us analyze if the symmetry is maintained also during spike emission,
in the usual case $v_1$ will reach $\infty$, then it will be reset
to $-\infty$ and a constant value $J_\delta$ will be added to all the other
membrane potentials. The membrane potential $w_1(t)$ reaching $\infty$ is equivalent 
to $v_N(-t)$ reaching $-\infty$.
Backwards in time the reset and coupling consists of setting  $v_N$ to $\infty$
and subtracting $J_\delta$ from the other variables. Due to the minus sign in the
definition of $\mathbf{w}(t)$ this means that $w_1$ is reset from $+\infty$ to $-\infty$
and the other variables are incremented by $J_\delta$. Hence $\mathbf{w}(t)$ is a solution
and \eqref{QIF} has time reversal symmetry.

We also show that the splay state is transformed to itself by the time reversal.
A splay state is a solution $\mathbf{v}(t)$  characterized by
the following properties
\begin{equation}
v_j(t+T)=v_{j+1}(t), \qquad  v_j(t+N T) = v_j(t) \qquad j=1,\ldots , N.  
\end{equation}
Note that if $\mathbf{w}(t)$ is the time reversal of
$\mathbf{v}(t)$ then $w_j(t)=-v_{N-j+1}(-t)$, $j=1, \ldots , N$.
We now make the following computation: 
\begin{align}
\label{eq-trsplay}
\begin{split} 
w_j(t+T)&=-v_{N-j+1}(-t-T)\\
&=-v_{N-j}(-t-T+T)\\ &=-v_{N-(j+1)-1}(-t)=w_{j+1}(t)
\quad .
\end{split} 
\end{align} 
It follows that $\mathbf{w}(t)$ is also a splay state. Moreover,
by choosing the phase, we can set $v_1(0)=0$, which implies that
$v_1(0)=w_N(0)$, or $v_1(0)=w_1((N-1)T)$. Therefore $\mathbf{w}(t)$ must be a
phase shifted version of $\mathbf{v}(t)$.

We now use the following well known result \cite{Lamb_1998}: 
\begin{thm}
\label{th-tr} 
Let
\begin{equation}
\label{eq-ode1} 
\dot{\mathbf{x}}= F(\mathbf{x}),  \qquad \mathbf{x}\in {\cal R}^N 
\end{equation} 
be an
ordinary differential equation and $R$ a matrix. Suppose that \eqref{eq-ode1} has
a time reversal symmetry defined as follows: if $\mathbf{x}(t)$ is a solution of
\eqref{eq-ode1} then $\mathbf{y}(t)=-R\mathbf{x}(-t)$ is also a solution. Suppose also that
\eqref{eq-ode1} has a periodic solution $\mathbf{x}_0(t)$ such that $-R\mathbf{x}_0(-t)=\mathbf{x}_0(t+T)$,
for some $T$. Then all the Floquet multipliers of $\mathbf{x}_0(t)$ are on the unit
circle. 
\end{thm}

\begin{figure}[htbp] 
\begin{center}
\includegraphics[draft=false,clip=true,height=0.32\textwidth]{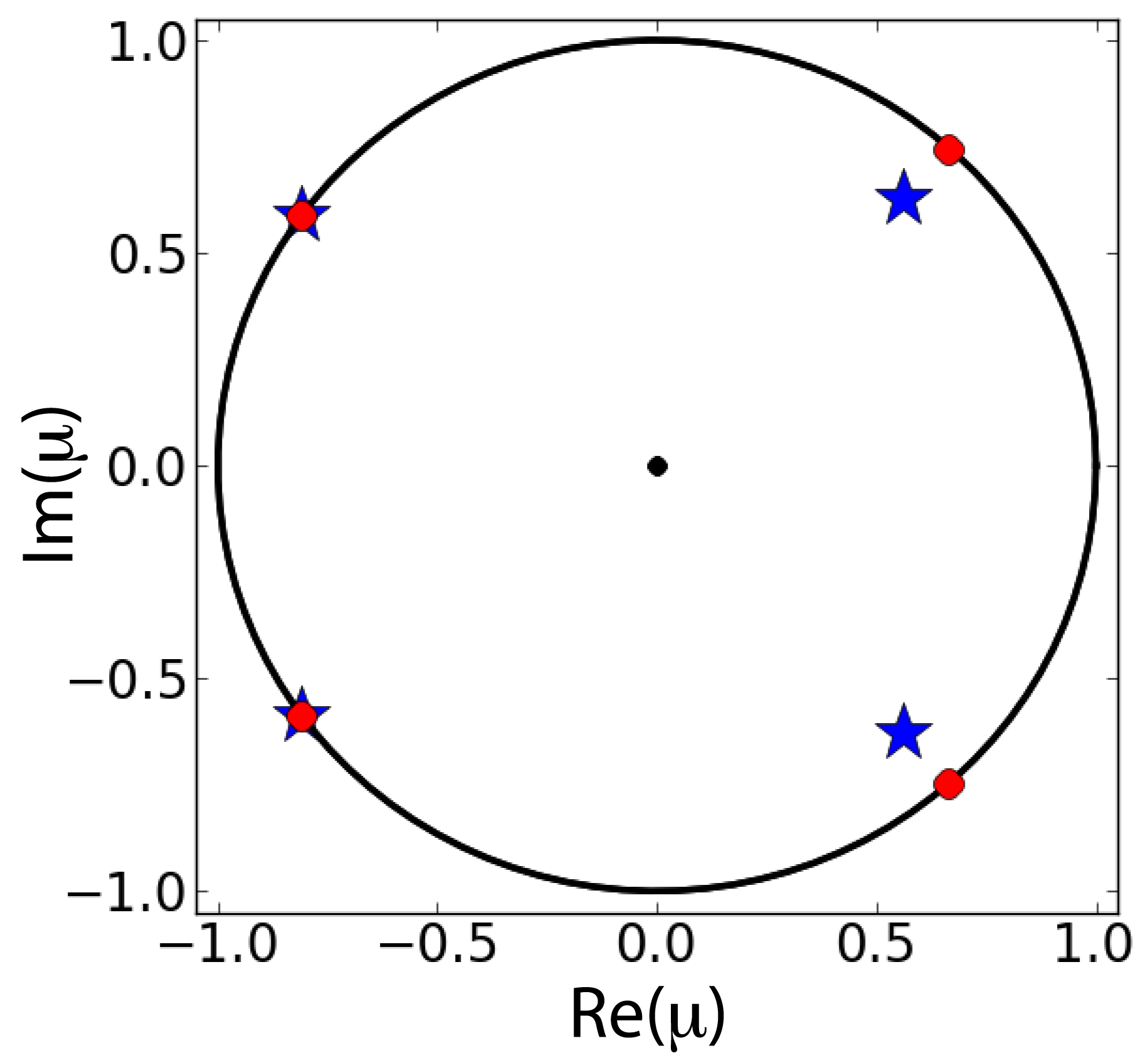}
\end{center}
\caption{Floquet multipliers for splay state with different PSPs:
namely, blue stars refer to step functions with $J=10$, 
and red circles to $\delta$-functions. The coupling
weight is the same in the two cases, $G=180$ ms.
} 
\label{fig:delta_comp}
\end{figure}

It follows from Theorem
\ref{th-tr} that the splay phase solution has all its Floquet 
multipliers on the unit circle (as shown in Fig. \ref{fig:delta_comp}).
In particular, in Fig. \ref{fig:delta_comp} we report the Floquet
multipliers for two different shape of PSP, but maintaining the the same coupling 
weight $G$, we observe that the multipliers which were inside the unit circle attain
modulus one by passing continuously from step to $\delta$-pulses.

\section{Continuous Family of Periodic Solutions}
We want to show that the $N-3$ directions of neutral stability for the splay state 
are not only local but also global.  We have verified this issue numerically,
by perturbing randomly the splay state $\tilde{\mathbf{x}}$ and by following
the system dynamics,  with the aid of the general event-driven map discussed in \S3A, until its convergence to some stationary state.
In particular, the initial conditions for these simulations have
been generated as follows
\begin{equation}
\mathbf{x}=\tilde{\mathbf{x}}+\sigma \mathbf{\mathcal{N}}
\qquad ,
\end{equation}
where $\tilde{\mathbf{x}}$ identifies the splay state,
$\mathcal{N}$ is a $N$-dimensional random vector whose components are 
$\delta$-correlated with zero average and Gaussian distributed
with unitary standard deviation, and the noise amplitude is $\sigma=0.1$.
By following the time evolution for a
sufficiently long time span (typically, of order of $100\cdot N$ spikes),
we always observe that these initial conditions converge to periodic orbits or to the quiescent state $\mathbf{x}=\{-1,...,-1\}$.
This has been verified for system
size up to $N=1,000$ and by considering up to $10,000$
different initial conditions for each $N$. 

\begin{figure}[htbp]  
\begin{center}
\includegraphics[draft=false,clip=true,height=0.7\textwidth]{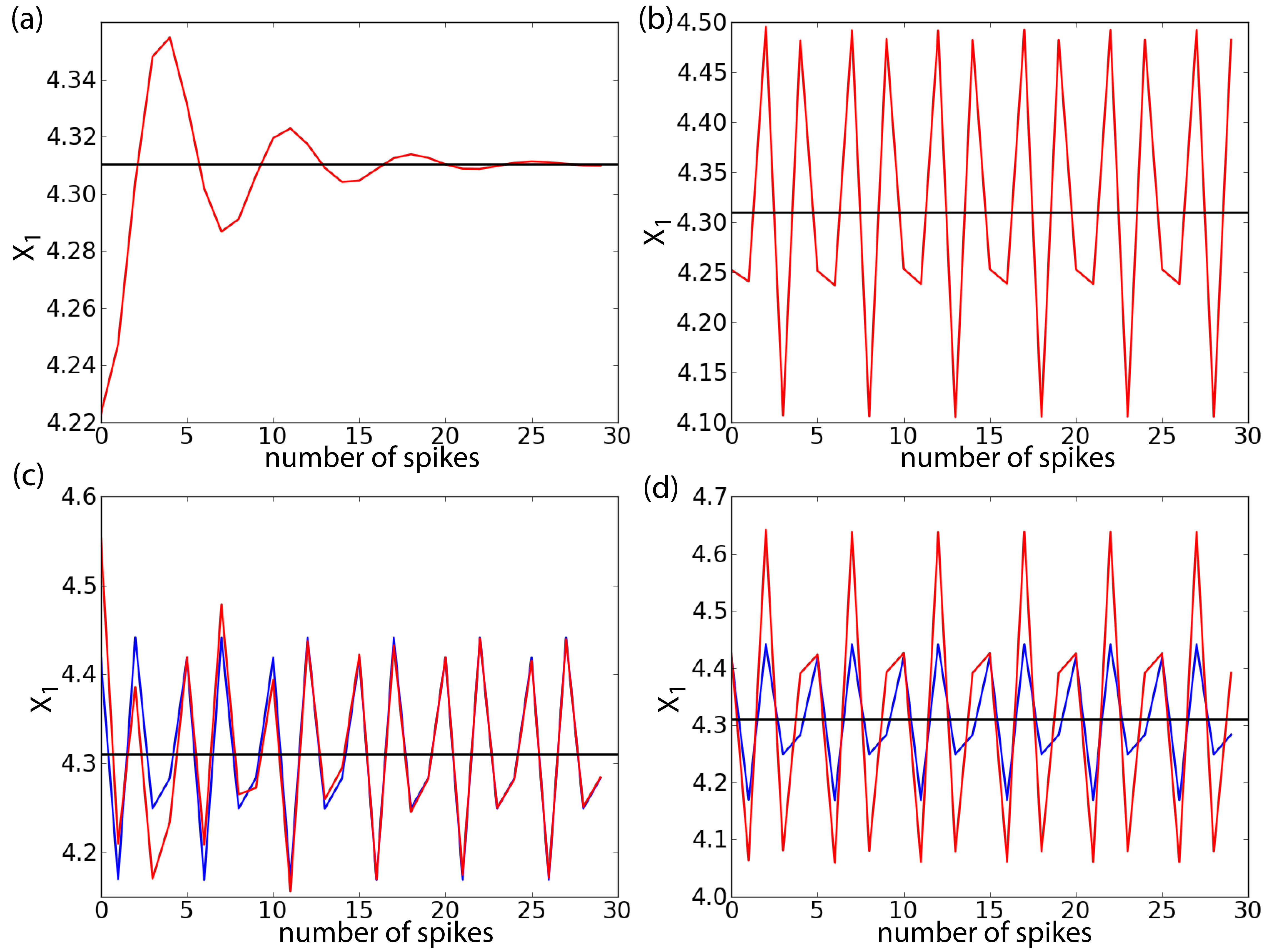}
\end{center}
\caption{Examples of trajectories (red lines) emerging from the perturbation of the splay state (black lines) or of a periodic state (blue lines). 
Only the voltage variable $x_1$ is reported here as a function of the index labeling the sequence of successive firings. Perturbation of the splay state: (a) along the directions of stability, the system converges to the splay state; (b) along the directions of neutral stability, the system is set in a periodic state. Perturbation of a periodic state: (c) along the directions of stability, the system converges to the periodic orbit; (d) along the directions of neutral stability, the system is set in a new periodic orbit. 
The system parameters are $N=5$, $J=15$, and $T_s=3$ ms, and $\sigma=0.2$.} 
\label{fig:FiguraD}   
\end{figure}

Furthermore we observe that the final state is an orbit with periodicity $\chi=N$
if $N>4$ and periodicity $\chi=2$ if $N=4$ (Fig. \ref{fig:FiguraD}). For
$N=3$ the final state is always the splay state. 
These solutions are characterized by neurons firing periodically
with the same period, but with time intervals among successive firing 
which are not constant, like for the splay state. Please notice that, in the
event-driven map context, the splay state amounts to a fixed point of the dynamics.

All the the periodic orbits we found lie on the $(N-3)$-manifold associated
to the neutrally stable directions of the splay state in the event-driven map formulation, that can be obtained by eq. \eqref{fin2}. 
We can affirm this, since on one hand we have verified that by perturbing the splay state along the stable directions we end up to the splay state itself (as shown in Fig. \ref{fig:FiguraD}(a)), while while by perturbing
along the
neutrally stable directions we always end up in one of these many periodic orbits (as shown in Fig. \ref{fig:FiguraD}(b)). 
On the other hand, by perturbing one of these orbits along the stable directions of the
splay state the perturbed system converges to the same orbit (see Fig. \ref{fig:FiguraD}(c)), while by perturbing
along the neutrally stable directions the system ends up on
a different periodic orbit (see Fig. \ref{fig:FiguraD}(d)). Therefore these periodic orbits are also
neutrally stable and share the same neutrally stable manifold of the splay state.

The existence of this manifold made of a continuous family 
of periodic solutions has been previously reported for Josephson
arrays \cite{Tsang_1992,Golomb_1992} and Watanabe and Strogatz discussed
the generality of this issue, reporting a ``heuristic'' argument to
support the existence of this manifold for generic fully coupled oscillator
networks with sinusoidal coupling \cite{watanabe}.

\begin{figure}[htbp]  
\begin{center}
\includegraphics[draft=false,clip=true,height=0.6\textwidth]{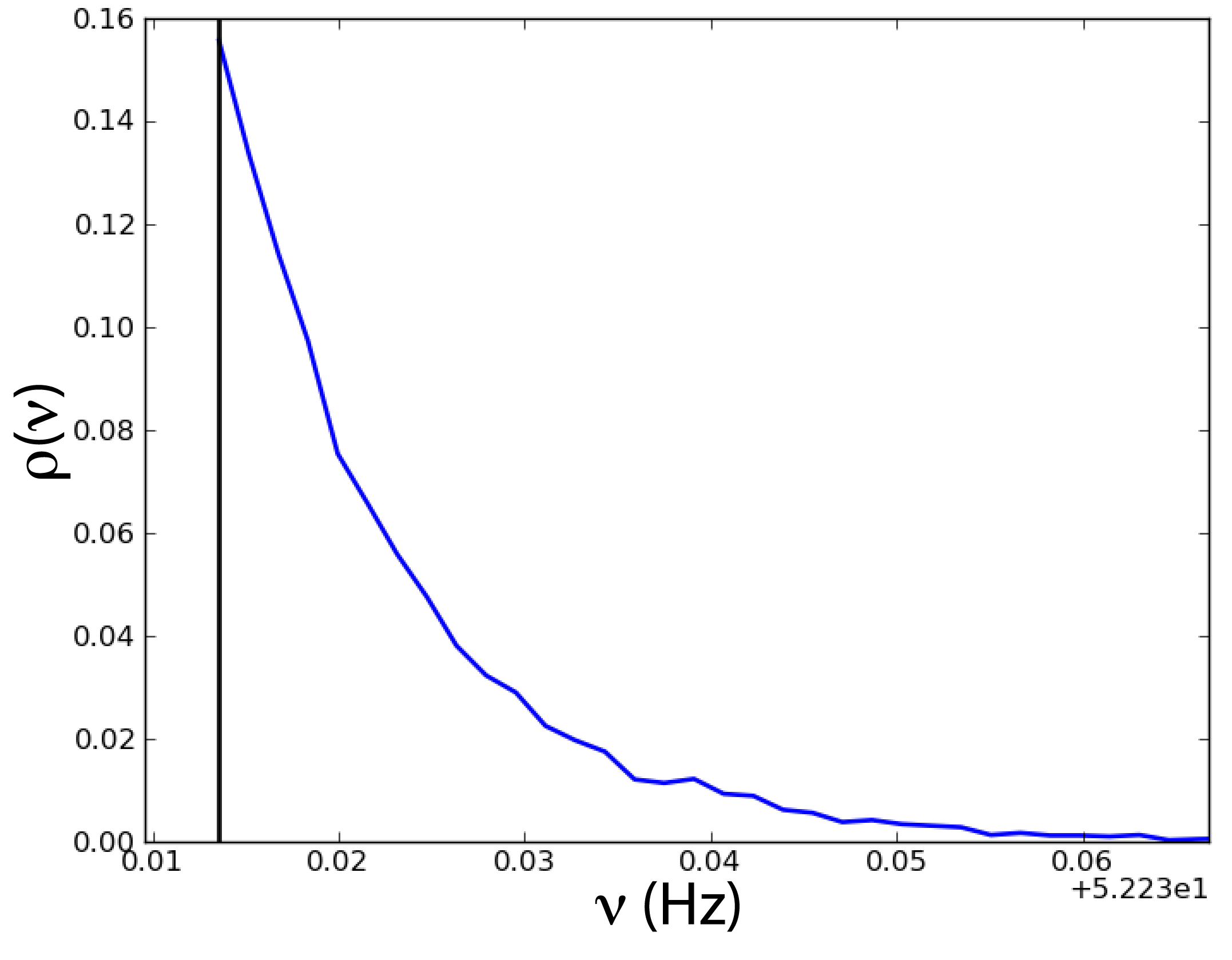}
\end{center}
\caption{Probability distribution of the single neuron firing rate $\nu$ for the
periodic solutions and the splay state (vertical black line). 
The $N_t$ initial condition are generated by randomly perturbing the splay state
along the directions of neutral stability, the perturbations were gaussian
distributed with zero average.The model parameter are the same as in the previous figure. 
and $N_t=10,000$.} \label{fig:FiguraE}  
\end{figure}

As a last point we have evaluated for the splay state and several 
 periodic orbits (namely $N_t=10,000$) the single neuron firing rate $\nu$. The distribution
of these rates is reported in Fig. \ref{fig:FiguraE}, revealing that 
the splay state is characterized by the minimal firing rate with respect to the ones found for the associated family of periodic orbits.

\section{Conclusions}
In this paper we showed analytically that finite-size all-to-all pulse-coupled excitatory networks  of excitable neurons admit marginally stable persistent splay states. We obtained analytical information about the stable firing rates of these sustained activities. 
Since the firing rate of persistent states  is an electrophysiologically measurable quantity in a working memory tasks, these results can provide insights for working memory models.
We further obtained results on the splay state stability that can help in choosing 
the correct parameters required for biologically relevant working memory models. Our results also give an analytical basis to previous observations in models that the stable sustained neural activity is asynchronous \cite{laingchow, compte}.

We developed event-driven map methods to analyze the network dynamics and found an analytical expression for the Floquet
spectra associated to the splay state for step pulses and $\delta$-pulses.
In the case of $M$ overlapping synaptic step pulses our analysis has revealed that for a
correct treatment of the linear stability analysis, the evolution of 
$M$ additional variables, corresponding 
to the last $M$ firing events, should be taken in account.

Our analysis, extending previous results for
systems with sinusoidal coupling \cite{watanabe}, revealed that
the splay state is marginally stable for finite size networks
with $N-2$ neutral directions, which reduce to $N-3$ in the
event-driven map formulation. We also reported 
a rigorous proof for non overlapping step pulses. We further identified a continuous family of periodic solutions surrounding the splay state. Their peculiarity is that these periodic states have exactly the same neutral stability directions as the splay state.

Our results leave several open questions, in particular we need
 to proof that at least one of the splay states is Lyapunov stable, when they exist.
It would be also of interest to extend the
rigorous results reported in \S C to overlapping
PSPs. Furthermore, since the stable persistently active solutions
 of our network have a specific spiking structure, splay or families
  of periodic solutions, it would be interesting to identify the structure
   of the unstable states that form the separatices between this sustained activity and the ground state.
Finally we should explain why all the marginally stable states are periodic.

\section*{Acknowledgments}
We thank Adrien Wohrer for constructive suggestions. MD was partially supported by MESR (France), MK by a grant from the city of Paris during his stay in France, and BSG by CNRS, ANR-Blanc Grant Dopanic, CNRS Neuro IC grant, Neuropole Ile de France, Ecole de Neuroscience de Paris collaborative grant and LABEX Institut des Etudes Cognitives.
AT acknowledges the Villum Foundation (under the 
VELUX Visiting Professor Programme 2011/12)
and the Joint Italian-Israeli Laboratory on Neuroscience,
funded by the Italian Ministry of Foreign Affairs, for partial support.

\appendix 

\section{Explicit solution of the splay state firing rate for small network sizes}
In this Appendix we will show how
it is possible to obtain explicitly the firing rate $\nu$ of the splay state, for 
$N=2,3,4$, and for $M=0$ (namely for $J<J_1$). 

\subsection{Step pulses}
Eq. \eqref{map2} can be rewritten in the following way
\begin{equation}
\label{F1_d} 
y^{(n)}_i= \frac{-(1-\gamma)+(1+\gamma)x_i^{(n)}}{(1+\gamma)-(1-\gamma)x^{(n)}_i} 
\end{equation} 
where we have made use of the variable
\begin{equation}
\label{gamma} 
\gamma= \exp(-2 T_1/\tau)
\qquad ;
\end{equation}
Since in the present case  $0 < T_s <T$ the values of $\gamma$ are bounded between 0 and 1.

By employing \eqref{map1}, \eqref{F1_d}, and \eqref{F3}, the coefficients
 of the event-driven map \eqref{Fm} can be rewritten as
\begin{equation}
\label{coef1} 
\begin{array}{l} 
a_0=-(1-\gamma)+(1+\gamma)(J-1)\beta_1(T_s)\\ 
a_1=(1+\gamma)+(1-\gamma)\beta_1(T_s) \\ a_2=(1+\gamma)-
(1-\gamma)(J-1)\beta_1(T_s)  \\  a_3=-(1-\gamma)-(1+\gamma)\beta_1(T_s)
\end{array}  
\end{equation}

The firing rate can be obtained in an explicit form by inverting \eqref{gamma},
namely
\begin{equation}
\label{firing_rate} 
\nu=\frac{1}{NT}=\frac{1}{N}\frac{1}{T_s-\frac{\tau}{2}\ln\left( \gamma(N,J,T_s)\right)}
\end{equation}
Once fixed the network parameters,
an admissible solutions for $\gamma \in [0;1]$ amounts 
finding a splay state solution with a frequency given by \eqref{firing_rate}.

Given an admissible $\gamma$ value, 
the membrane potentials corresponding to the splay state can be found 
by iterating the map \eqref{Fm} starting from the boundary condition $\tilde{x}(N)=-\infty$
corresponding to the reset value, namely
\begin{equation}
\label{V_values} 
\left( \begin{array}{c}  \tilde{x}_{N-1}\\
\tilde{x}_{N-2}\\ \vdots \\ \tilde{x}_2\\ \tilde{x}_1\\ \tilde{x}_0\\
\end{array} \right)=  \left(\begin{array}{c}  a_1/a_3\\ (a_0 a_3
+a^2_1)/(a_3(a_1+a_2)) \\ \vdots  \\ -(a_0a_3+a^2_2)/(a_3(a_1+a_2))\\ 
-a_2/a_3\\  \infty\\ \end{array} \right) 
\end{equation}

We can finally determine $\nu$ analytically for $N=2,3,4$.

\begin{itemize}

\item In the case of a couple of neurons, $N=2$, we should impose
$\tilde{x}_1=\tilde{x}_{N-1}$ and thus we have $a_1+a_2=0$. Solving this
equation for $\gamma$ we obtain: 

\begin{equation}
\label{gamma2} 
\gamma=\frac{(J-2)\beta_1(T_s)-2}{(J-2)\beta_1(T_s)+2}
\end{equation}

in this case we have a unique stable branch of solutions, as shown in Fig. \ref{fig:fvsg} 
Furthermore, the minimal reachable frequency is zero and it is achieved for $\gamma=0$,
when $J$ and $T_s$  satisfy the equation $(J-2)\beta_1(T_s)=2$.

\item For $N=3$ we have $\tilde{x}_1=\tilde{x}_{N-1}$, using the values in
\eqref{V_values} we obtain:
\begin{equation}
a_0 a_3 +a^2_1+a^2_2+a_1 a_2=0
\end{equation}
and then, we can reorder this equation as a second order equation for
$\gamma$:
\begin{equation}
\label{gamma_pol_3}
[(J-2)\beta_1(T_s)+2]^2\gamma^2-2[(J^2-2J+2)\beta^2_1(T_s)-2]\gamma+[(J-2)\beta_1(T_s)-2]^2=0
 \end{equation}
this equation admits the following two solutions
\begin{equation}
\label{gamma_3}
\begin{array}{l}
\gamma_{1,2}=\{[(J^2-2J+2)\beta^2_1(T_s)-2]\pm \\
\\
\pm \sqrt{[(J^2-2J+2)\beta^2_1(T_s)-2]^2-[(J-2)^2\beta^2_1(T_s)-4]^2}\}\cdot\\
\\
\cdot \{[(J-2)\beta^2_1(T_s)+2]^2\}^{-1}
\end{array}
\end{equation}
$\gamma_1$ (resp. $\gamma_2$) is associated to the upper stable (resp. lower unstable) branch
reported in Fig. \ref{fig:fvsg}. In this case the upper branch is bounded away from the zero
frequency, and the minimal frequency is attained for $\gamma_1 = \gamma_2$, when $J$ and
$T_s$ satisfy $(J^2-3J+3)\beta^2_1(T_s)=3$. The zero frequency is instead reachable on the lower
branch for $\gamma=0$ as shown in Fig. \ref{fig:fvsg}.

\item 
If $N=4$ then $\tilde{x}_2=\tilde{x}_{N-2}$ and the coefficients should satisfy the
following equation 
\begin{equation}
2a_0 a_3 +a^2_1+a^2_2=0 \qquad .
\end{equation}
Similarly to the case $N=3$ we obtain
a quadratic equation for the parameter $\gamma$, namely
\begin{equation}
\label{gamma_pol_4}
[(J-2)\beta^2_1(T_s)+2]^2\gamma^2-2[J^2\beta^2_1(T_s)]\gamma+[(J-2)\beta_1(T_s)-2]^2=0 
\quad ,
\end{equation}
also in this case we have 2 branches of solutions for the splay state frequencies 
parametrized by $\gamma_1$ and $\gamma_2$
\begin{equation}
\label{gamma_4} \gamma_{1,2}=\frac{J^2\beta^2_1(T_s)\pm
\sqrt{[J^2\beta^2_1(T_s)]^2-[(J-2)^2\beta^2_1(T_s)-4]^2}}{[(J-2)\beta^2_1(T_s)+2]^2}
\quad . 
\end{equation}
Also in this case the zero frequency is attainable on the unstable branch 
for $\gamma=0$ and the merging of stable and unstable branch occurs at a finite
frequency corresponding to a value of $J$
which is solution of $(J^2-2J+2)\beta^2_1(T_s)=2$.

\end{itemize}


\subsection{$\delta$-pulses}
We can rewrite the coefficients \eqref{coef22} of the map \eqref{Fm} for the case of $\delta$-pulses combining
\eqref{map2_d}, \eqref{F1_d} and \eqref{F3} as follows
\begin{equation}
\label{coef2} \begin{array}{l}   a_0=-(1-\gamma)
+(1+\gamma) J_{\delta}\\ 
a_1=(1+\gamma)\\
a_2 =(1+\gamma)-(1-\gamma) J_{\delta}  \\
a_3=-(1-\gamma)\end{array} 
\quad ,
\end{equation}
where $\gamma$ is given by  the expression \eqref{gamma} with $T_1= T$.
The firing rate for the splay state can be
obtained from the following expression
\begin{equation}
\label{firing_rate1} 
\nu=-\frac{1}{\frac{N\tau}{2}\ln\left(\gamma(N,G,\tau)\right)} 
\end{equation}

Let us now discuss of the existence of the splay state for for $N=2,3,4$:

\begin{itemize}

\item In the case of a couple of neurons, $N=2$, solving this equation for $\gamma$ 
one obtains:
\begin{equation}
\label{gamma_2} 
\gamma=\frac{J_\delta-2}{J_\delta+2} 
\quad ;
\end{equation}
like in the step pulses case one has only one branch and
the splay state exists for $J_\delta >2$ and the period diverges to infinite at $J_\delta = 2$.
\item If $N=3$
\begin{equation}
\label{gamma_3d} 
\gamma_{1,2}=\frac{J_\delta^2-2\pm
2\sqrt{J_\delta^2-3}}{(J_\delta+2)^2}
\qquad ; 
\end{equation}
now two branches are present and similarly to 
the step pulses the upper branch (corresponding to $\gamma_1$)
is stable while the other one is unstable. The branches exist for 
$ J_\delta >\sqrt{3}$ and they merge exactly for this coupling value.
 
\item For $N=4$
\begin{equation}
\label{gamma_4d} 
\gamma_{1,2}=\frac{J_\delta^2\pm
2\sqrt{2 J_\delta^2-4}}{(J_\delta +2)^2} 
\qquad ;
\end{equation}
also in this case the two branches are present above
a certain critical coupling given by
$ J_\delta = \sqrt{2}$.

\end{itemize}

\section{Analytic expression for the splay state in the infinite size limit}
In the limit of $N\rightarrow\infty$ it is possible to derive an analytic
expression for the membrane potentials associated to the splay state both for
step and $\delta$-pulses. In such a limit the mean input current $I$ can be
assumed to be constant, and it can be  easily obtained from  \eqref{HM1}, giving
$I=\pi^2\tau^2\nu^2+1$. Thus we can rewrite \eqref{QIF} as follows
\begin{equation}
\label{QIF_infinity} 
\tau\frac{dv}{dt}=v^2+ \pi^2\tau^2\nu^2
\end{equation}

\begin{figure}[htbp] \begin{center}
\includegraphics[draft=false,clip=true,height=0.6\textwidth]{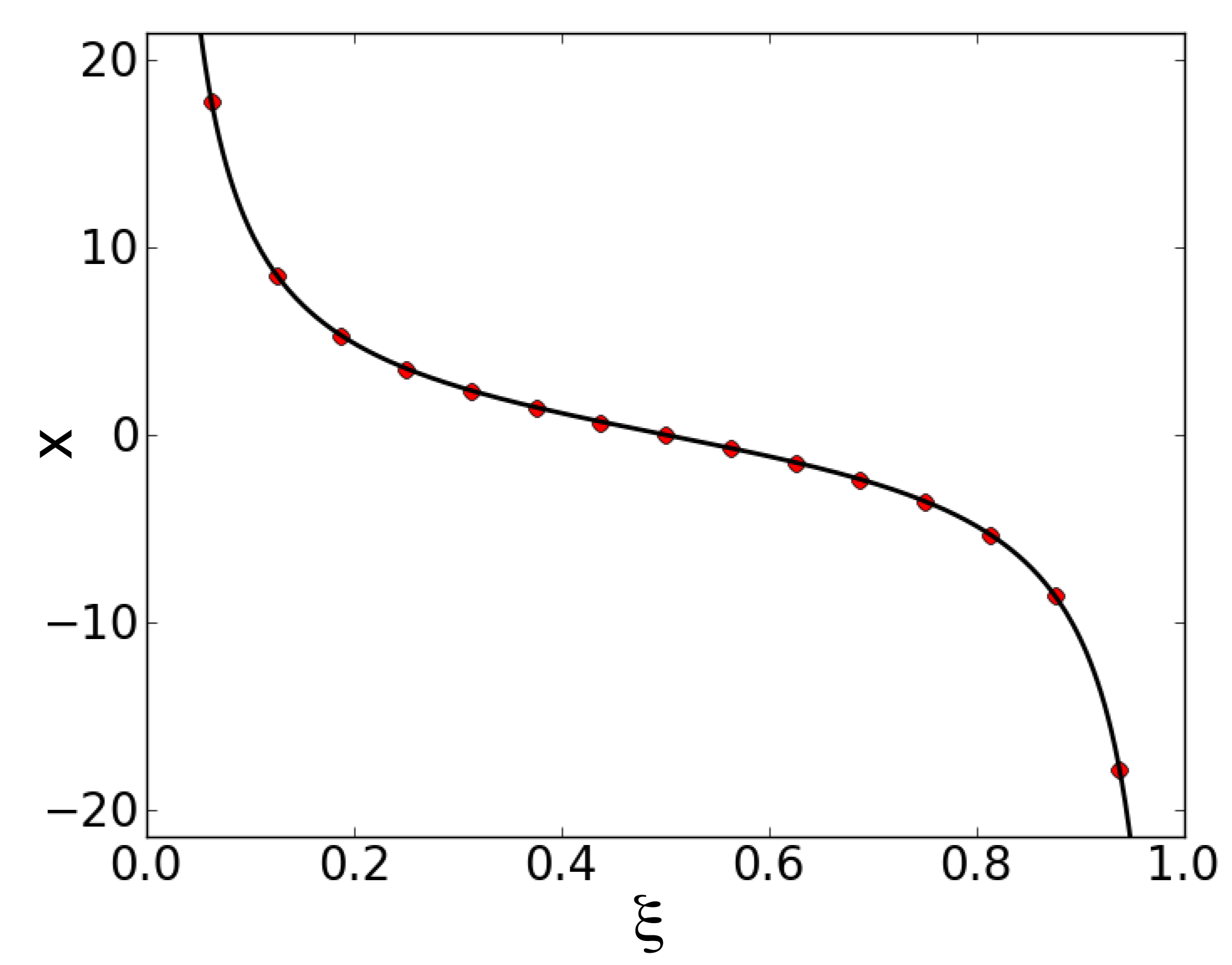}
\caption{Membrane potential values as a function of $\xi = i/N$ for a splay state.
The symbols refer to $N=16$, while the solid line to the continuous limit 
approximation. The data have been obtained for $J=15$, $T_s=1$ ms and
$\tau=20$ ms.} 
\label{fig:splay_voltage} \end{center}
\end{figure}

We can then integrate equation \eqref{QIF_infinity} between the reset value
$v=-\infty$ and  a generic time $t_i$:
\begin{equation}
\label{integral1} 
\int^{v(t_i)}_{-\infty} \frac{d v}{v^2 + \pi^2\tau^2\nu^2} = \int^{t_i}_{0} \frac{dt}{\tau} 
\end{equation}
the integration gives:
\begin{equation}
\label{integral2}
v(t_i)=-\pi\tau\nu\tan(\frac{\pi}{2}-\frac{\pi}{NT}t_i) 
\end{equation}
where for the splay state $\nu = 1/ (NT)$.
If we identify $t_i$ with the spike time of neuron $i$ in the network,
we will have that the splay state solution 
for the membrane potential of neuron $i$ is ${\tilde x}_i = - v(t_i)$, 
please notice that potential ${\tilde x}_i$ are ordered from the largest
to the smallest. Furthermore, since the spike times are equally spaced
for the splay solution as $t_i=i T$, $i=0,1,...N$ we can rewrite
\eqref{integral2} as
\begin{equation} 
\label{anal}
\tilde{x}_i=-v(t_i)=\frac{\pi\tau\nu}{\tan(\pi\frac{i}{N})}
\xrightarrow{N \to \infty} {\tilde x}(\xi) = \frac{\pi\tau\nu}{\tan(\pi \xi)}
\end{equation}
where $0 \le \xi \le 1$ is a continuous {\it spatial} variable.
As shown in Fig. \ref{fig:splay_voltage}, 
the expression obtained in the continuous limit compare reasonably well with
the numerically estimated finite size solutions already for $N=16$.

\section{Marginally stable directions of the splay states} 
In this appendix, we will analyze the stability of a splay state in the case
of non-overlapping step pulses. To perform this analysis, let us rewrite the QIF model \eqref{QIF} as follows
\begin{equation}
\label{theta}  
\tau \frac{d \theta_i}{dt}=I(t) +(I(t)-2) \cos \theta_i \qquad i=1,\dots,N 
\qquad ;
\end{equation} 
where we have performed the transformation
of variable $\theta_i =2 \tan^{-1}(v_i)$.
Therefore, the membrane potential is now represented by a a phase variable 
$\theta_i \in [-\pi;\pi]$, the spike is emitted
(and transmitted instantaneously to all the neurons in the network)
whenever $\theta_i$ reaches the threshold $\pi$ and then $\pi$ it is reset to $-\pi$.  The model in the formulation \eqref{theta} is termed $\theta$-neuron,
we will apply the Watanabe and Strogatz~\cite{watanabe} approach 
to this model to derive the Floquet spectrum for the splay state solution.

In order, to stress the peculiar PSPs we are considering, we rewrite
\eqref{theta} as follows 
\begin{equation}
\label{eq-theta}
\frac{d\theta_j}{dt}=(J\phi(\mathbf{\theta})-2)\cos(\theta_j)+J\phi(\mathbf{\theta}) 
\qquad\qquad j=1,\ldots N 
\end{equation}
with
$\phi(\theta)$ being the characteristic function of the interval
$[-\pi,\theta_{OFF}]$. The emission of a spikes occurs whenever the neuron $\min\{\theta_i\}=-\pi$
this amounts to an increase by one in the value of the function $\phi(\theta)$.
Furthermore,
when the pulse expires after a time $T_s$ the value of $\phi(\mathbf{\theta})$ will be decreased by one.
By assuming that no neuron will fire while the synapse is on (no overlapping PSPs), 
the PT will occur for a specific value of the phase variable  namely  $\min\{\theta_i\}=\theta_{OFF}$, for a value $\theta_{OFF}$, which can be determined as outlined in \S3A.

Let us now recall the approach devised by Watanabe and Strogatz~\cite{watanabe}
to show that each trajectory representing the dynamics
of a system of $N$ identical phase oscillators, whose evolution
is ruled by ODEs of the form
\begin{equation}
\label{eq-thetaws}
\frac{d\theta_j}{dt}=f(\mathbf{\theta})+g(\mathbf{\theta})\cos(\theta_j)+h(\mathbf{\theta})\sin(\theta_j) \qquad\qquad
j=1,\ldots N \quad, \end{equation}
is confined to a three dimensional subspace. 
The only requirement is that the functions $f$, $g$ and $h$ 
do not depend on the index $j$ of the considered oscillator.
In other words $f$, $g$ and $h$ are collective variables determined
by the network state.  Clearly our equation \eqref{eq-theta} satisfies this
condition. 

Watanabe and Strogatz introduce a transformation $Q_x:R^{N}\rightarrow R^{N+3}$
from variables $\{ \theta_j \}$ to variables $X \equiv (\Gamma,\Theta,\Psi,\{\psi_j\})$
defined implicitly by the set of equations
\begin{equation}
F(\theta_j,\Gamma,\Theta,\Psi,\psi_j)=0,\quad j=1,\ldots N \quad ;
\label{F}
\end{equation}
where
\begin{equation}
F=\tan\left(\frac{1}{2}(\theta_j-\Theta)\right)-\sqrt{\frac{1+\Gamma}{1-\Gamma}}
\tan\left(\frac{1}{2}(\psi_j-\Psi)\right). 
\end{equation}
Furthermore, they prove that an arbitrary solution of \eqref{eq-thetaws} can
be generated by the transformation $Q_x$ from a set of parameters $\{\psi_j\}$,
which remain constant in time, whenever the three collective variables
$\Gamma,\Theta,\Psi$ satisfy the following equations
\begin{align}
\label{eq-eqsws}
\begin{split} 
&\dot\Gamma =-(1-\Gamma^2)(g\sin\Theta - h\cos\Theta)\\
&\Gamma\dot\Theta=\Gamma f -g\cos\Theta -h\sin\Theta\\
&\Gamma\dot\Psi=\sqrt{1-\Gamma^2} (g\cos\Theta +h\sin\Theta)  \qquad ,
\end{split}
\end{align}
and obviously the other variables satisfy
\begin{equation}
\label{eq-eqsws2}
\dot \psi_j = 0  \qquad \forall j=1, \dots, N
\qquad .
\end{equation}

We prove the following proposition: 

\begin{proposition}  \label{prop-splay} Let us assume that \eqref{eq-theta}
admits a splay state solution and that this solution is Lyapunov stable. 
Then at least $N-2$ Floquet multipliers will lie on the unit circle.
\end{proposition}

Let us now recall the definition of the Floquet multipliers~\cite{levinson}
for a generic ODE of the form
\begin{equation}
\label{eq-ode} 
\dot{\mathbf{\theta}}= F_x(\mathbf{\theta}), \qquad \mathbf{\theta} \in R^N \quad,
\end{equation}
admitting a periodic solution $\mathbf{\theta}^s(t)$  with period $T_p$. 

The associated variational linear equation in the tangent space 
has the form:
\begin{equation}
\label{eq-varode} 
\delta \dot{\mathbf{\theta}}= DF_x(\mathbf{\theta}^s(t))\delta \mathbf{\theta}, \qquad \delta \mathbf{\theta} \in {\cal R}^N. 
\end{equation}
Equation \eqref{eq-varode} has (possibly complex) eigensolutions $\Phi(t)=e^{(\lambda+i\omega) t}\eta(t)$, with $\eta(t)$ periodic of period $T_p$, termed Floquet vectors. The complex numbers $\mu(T_p)=e^{(\lambda+i\omega) T_p}$ are the Floquet multipliers. They determine the stability of the periodic solution.

{\em Proof of \ref{prop-splay}}. 
We will prove that there exists an $N-2$ dimensional subspace of solutions of the variational equation
associated to \eqref{eq-theta} consisting of 
solutions that do not converge to the ${\bf 0}$ vector as $t\to \infty$ 
(except for the ${\bf 0}$ solution itself).  This, combined with Lyapunov stability, implies that there
must be $N-2$ Floquet multipliers on the unit circle.

Let $\mathbf{\theta}^s_0=\{\theta_{0,1}^s,\ldots ,\theta_{0,N}^s\}$ be a choice of initial
conditions corresponding to a splay state of period $T_p$. For simplicity and without
any loss of generality, we can assume that the phases
are ordered, i.e. $\theta_{0,1}^s>\theta_{0,2}^s>\ldots ,\theta_{0,N}^s$, and that
$\theta_{0,1}^s$ is close to $\pi$, i.e. the first neuron is just about to fire. 

Let us consider a solution $\mathbf{X}^s(t)$ of \eqref{eq-eqsws} and \eqref{eq-eqsws2} with 
initial condition 
\begin{equation}
\mathbf{X}^s_0=\{0,\frac{\pi}{2},\frac{\pi}{2},\theta_{0,1}^s,\ldots , \theta_{0,N}^s\} \quad ;
\end{equation}
where it is evident that $\theta^s(t)=T_x(\mathbf{X}^s(t))$ since $T_x(\mathbf{X}^s(0))=\theta^s(0)$.
Furthermore, we perturb the initial condition with a perturbation of the form
\begin{equation}
\Delta\psi=(\Delta\mathbf{\psi}_1,\ldots,\Delta\psi_{N-2},0,0) \qquad,
\end{equation}
 in the following way
\begin{equation}
\mathbf{X}^s_{\Delta\psi}(0)= \mathbf{X}^s_0 + \Delta\psi =
\{0,\frac{\pi}{2},\frac{\pi}{2},\theta_{0,1}^s+\Delta\psi_1,\ldots
,  \theta^s_{0,N-2}+\Delta\psi_{N-2},\theta_{0,N-1}^s,\theta_{0,N}^s\} \qquad. 
\end{equation}
and we obtain the perturbed solutions $\mathbf{X}^s_{\Delta\psi}(t)$
at time $t$ by integrating \eqref{eq-eqsws} and \eqref{eq-eqsws2},
while the the corresponding solution of \eqref{eq-thetaws} is given by
$\theta^s_{\Delta\psi}(t)=T_x(\mathbf{X}^s_{\Delta\psi}(t))$. 

Let us denote the value of the perturbed orbit at integer multiples
$k$ of the period $T_p$ as follows  
\begin{equation}
\mathbf{X}^s_{\Delta\psi,k} =
(\Gamma(kT_p),\Theta(kT_p),\psi(kT_p),\theta_1^s
+\Delta\psi_1,\ldots ,
\theta_{N-2}^s+\Delta\psi_{N-2},\theta_{N-1}^s,\theta_N^s) \quad,
\end{equation}
and $\theta^s_{\Delta\psi,k}=T_x(\mathbf{X}^s_{\Delta\psi,k})$. 

We will show that there exists a real positive constant $L$ such that, for every value of $k$, 
\begin{equation}
\label{eq-est}
\|\theta^s_{\Delta\psi,k}-\theta_0^s\|\ge L\|\Delta\mathbf{\psi}\|. 
\end{equation}

By assuming that the perturbation is sufficiently small,
i.e. $||\mathbf{X}^s_{\Delta\psi,k}-\mathbf{X}_0^s|| \ll 1$, we can approximate
the evolution of the perturbed orbit in proximity of the unperturbed
one, with the corresponding linearized dynamics, namely
\begin{equation}
T_x(\mathbf{X}^s_{\Delta\psi,k})-T_x(\mathbf{X}^s_0) \approx
DT_x(\mathbf{X}_0^s) (\mathbf{X}^s_{\Delta\psi,k}-\mathbf{X}_0^s)
\qquad .
\end{equation}
In order to write the Jacobian $DT_x(\mathbf{X}_0^s)$, we need to estimate
the following derivatives, which can be obtained by implicit differentiation
of \eqref{F}
\begin{equation}
\label{eq-derivs}
\frac{\partial\theta_j}{\partial\Theta}=1,\; 
\frac{\partial\theta_j}{\partial\Psi}=-1,\;
\frac{\partial\theta_j}{\partial\gamma}=-\cos\theta_j^s,\;
\frac{\partial\theta_j}{\partial\psi_k}=\delta_{jk}.   
 \end{equation}
where $\delta_{jk}$ is the Kronecker delta.

Let ${\bf V}^0\in {\cal R}^{N}$ be a vector with an unitary norm spanning a $N-2$ dimensional subspace
and let assume that  $\Delta{\bf \psi}=\sigma \mathbf{V}^0$ with $0<\sigma\le 1$. We will prove that
\begin{equation}
\label{eq-finest}
\|\frac{1}{\sigma} DT_x(\mathbf{X}_0^s)(\mathbf{X}^s_{\Delta\psi,k}-\mathbf{X}_0^s)\|\ge L>0 \qquad, 
\end{equation}
for some real constant $L$ independent of $\sigma$ and $k$. 

By employing \eqref{eq-derivs} the following expression
can be derived
\begin{equation}
\label{eq-bigvect} \frac{1}{\sigma}
DT_x(\mathbf{X}_0^s)(\mathbf{X}^s_{\Delta\psi,k}-\mathbf{X}_0^s)=\left(\begin{array}{l}
Z_1(k)/\sigma+V^0_1\\
Z_2(k)/\sigma +V^0_2
\\ \quad \cdot \\ \quad\cdot \\
\quad\cdot \\ 
Z_{N-2}(k)/\sigma+V^0_{N-2}\\
Z_{N-1}(k)/\sigma\\ 
Z_{N}(k)/\sigma
\end{array}\right ) 
\end{equation}
where for brevity and clarity we set $Z_j (k)= v_k - z_k - \cos\theta_{j}^s v_k$
once redefined $v_k=\Gamma(kT_p)$,
$w_k=\Theta(kT_p)-\frac{\pi}{2}$, and $z_k=\Psi(kT_p)-\frac{\pi}{2}$.
It is clear, due to their definition,  that the components of
the vector ${\bf Z}(k) = \{Z_j(k)\}$ are not linearly independent
and in particular that they span a 2-dimensional subspace.

As a first step, the validity of the following inequality,
$\forall k$ and for any sufficiently small $\sigma$, is discussed 
\begin{equation}
\label{eq-compest}
|Z_j(k)|/\sigma = |(w_k-z_k-\cos\theta_{j}^sv_k)|/\sigma\ge L\quad\mbox{for $j=N$ or $j=N-1$ } 
\qquad .
\end{equation}

We consider two possible cases. 
In the first case, the inequality \eqref{eq-compest} holds, therefore
\eqref{eq-finest} is satisfied since the length
of any vector is bigger than the absolute value of one of its components, 
thus implying that the modulus of the l.h.s. of \eqref{eq-bigvect} would be greater than $L$
for any $k$ value.

In the second case, we assume that \eqref{eq-compest} does not hold
uniformly in $k$ for $j=N$, and $j=N-1$, in other words the components
$|Z_{N-1}(k)|/\sigma$ and $|Z_N(k)|/\sigma$  should converge to $0$
for $k \to \infty$ and $\sigma \to \infty$. 
Furthermore, since for $N>3$ $\cos\theta_N^s \ne \cos\theta_{N-1}^s$,
each component $Z_j$ with $j=1, \ldots, N-2$ can be written as a linear
combination of $Z_{N-1}$ and $Z_N$. This implies that each element
$|Z_{N-1}(k)|/\sigma$ remains arbitrarily small $\forall j$
even for arbitrarily large (resp. small) $k$ (resp. $\sigma$).
Now each component in the r.h.s. of \eqref{eq-bigvect} will have
the form $Z_j/\sigma + V^0_j$ for $j=1, \ldots, N-2$, where the first
quantity is arbitrarily small, but by construction
the vector ${\bf V}^0$ has an unitary modulus, thus also in this second
case  \eqref{eq-finest} is satisfied for any $k$.

From the previous results it follows that
the vector function 
\begin{equation}
 {\bf V}(t)=\frac{d}{d\sigma}
\theta_{\Delta\Psi}(t) |_{\sigma=0} 
\end{equation}
is a solution of the variational equation \eqref{eq-varode}
which does not converge to $0$ as $t\to\infty$. 
Since \eqref{eq-varode} is a system of linear equations, a vector space
of initial conditions gives rise to a vector space of solutions.
Since ${\bf V}^0$ spans a
$N-2$ dimensional vector space, which we denote by $LV$, 
our construction give a $N-2$ dimensional vector
space of solutions of \eqref{eq-varode}, which we denote by ${\cal LV}$.

As mentioned above the Floquet vectors are solutions of \eqref{eq-varode}
 of the form $\mu(t)\eta(t)$, with $\eta(t)$ periodic of period $T_p$ and $\mu(T_p)$
  the corresponding Floquet multipliers. Since we assumed that the examined periodic
   orbit (i.e. the splay state) is Lyapunov stable, the multipliers $\mu(T_p)$ must be
    either on the unit circle or inside the unit circle. Without loss of generality, let us
     assume that at least two multipliers are inside the unit circle, otherwise the theorem would be automatically true.

Let denote by $LW$ the vector space spanned by the initial conditions of the
two Floquet eigenvectors associated to the two multipliers which
lie inside the unit circle and
let ${\cal LW}$ be the corresponding vector subspace of solutions
(spanned by the two Floquet eigenvectors). 
Since all non-zero solutions in ${\cal LW}$ converge to $0$ as $t\to \infty$
it follows that the intersection of $LW$ and $LV$ consists of the zero vector.
Therefore, we can formally decompose any of the remaining $N-2$ Floquet vectors
at initial time $t=0$ in two vectors, namely $\eta(0) = {\bf W}_1(0) + {\bf V}_1(0)$
where ${\bf W}_1(0) \in LW$ and ${\bf V}_1(0) \in LV$. By linearity, if 
${\bf W}_1(t)$ and ${\bf V}_1(t)$ are the solutions of \eqref{eq-varode} with initial conditions
${\bf W}_1(0)$  and ${\bf V}_1(0)$, it follows that 
$\eta (t)={\bf W}_1(t)+{\bf V}_1(t)$. If ${\bf V}_1(0)\neq {\bf 0}$ then $\eta(t) \not\in {\cal LW}$, moreover
$\eta(t)$ does not converge to $0$ as $t\to\infty$ since
${\bf W}_1(t)$ does, while $V_1(t)$ does not.
Therefore the corresponding Floquet multiplier can be only on the unit circle, due to our
previous assumptions. Finally we have demonstrated that $N-2$ Floquet multipliers are on the unit 
circle and 2 are inside the unit circle.

 \end{document}